\documentclass[final,5p,times,twocolumn]{elsarticle}

\usepackage{amsmath, amssymb, amsfonts}
\usepackage{bm}
\usepackage{graphicx}
\usepackage{booktabs}
\usepackage{enumitem}
\usepackage{xcolor}
\usepackage{mathtools}
\usepackage{hyperref}
\usepackage{cleveref}

\hypersetup{
  colorlinks=true,
  linkcolor=black,
  citecolor=black,
  urlcolor=black
}


\journal{Communications in Nonlinear Science and Numerical Simulation}

\newcommand{\R}{\mathbb{R}}

\newcommand{\norm}[1]{\left\lVert #1 \right\rVert}
\newcommand{\ip}[2]{\left\langle #1,#2 \right\rangle}
\newcommand{\tr}{\mathrm{tr}}
\newcommand{\corr}{\mathrm{corr}}

\newcommand{\x}{\bm{x}}

\newcommand{\Fhat}{\widehat{\bm{F}}}

\newcommand{\Gam}{\bm{\Gamma}}
\newcommand{\rhat}{\widehat{\bm r}}
\newcommand{\nhat}{\widehat{\bm n}}

\setlist[itemize]{topsep=3pt,itemsep=2pt,parsep=0pt,leftmargin=1.4em}
\setlist[enumerate]{topsep=3pt,itemsep=2pt,parsep=0pt,leftmargin=1.6em}

\begin{document}

\begin{frontmatter}

\title{Inferring Non-Normal Amplification Geometry from Multivariate Time Series}

\author[label1]{V.R. Saiprasad\fnref{equalcontrib}}
\author[label1]{V. Troude\fnref{equalcontrib}}
\author[label1]{D. Sornette\corref{cor1}}
\ead{dsornette@ethz.ch}
\cortext[cor1]{Corresponding author}
\fntext[equalcontrib]{These two authors contributed equally to this work.}

%\affiliation[label1]{  organization={Institute of Risk Analysis, Prediction and Management (Risks-X), Academy for Advanced Interdisciplinary Sciences, Southern University of Science and Technology},  city={Shenzhen},  country={China}}

\address[label1]{Institute of Risk Analysis, Prediction and Management (Risks-X),
Academy for Advanced Interdisciplinary Sciences,
Southern University of Science and Technology, Shenzhen, China}

 \begin{abstract} 
 Across hydrodynamics, ecology, neuroscience, network dynamics, non-Hermitian physics, and socio-economic systems, asymptotically stable dynamics can nevertheless exhibit large transient amplifications that remain invisible to analyses based solely on the eigenvalue spectrum. The mechanism is geometric rather than spectral: perturbations entering along some directions can be transiently expressed along distinct response directions, allowing asymptotic decay to coexist with strong transient or noise-driven amplification.
We introduce non-normal directional response inference, a data-driven method for detecting this geometry from multivariate time series when the governing operator is unknown. The method estimates a local linear operator from sliding windows of the multivariate recording, projects it onto the dominant two-dimensional input-response subspace, and summarizes the reduced dynamics by three diagnostics: the reduced eigenvalue splitting $\Delta$, the reduced eigenvector non-orthogonality $K$, and the scale-free ratio $R=K/K_c(\Delta)$, where \(K_c(\Delta)\) is the two-dimensional transient-amplification threshold.
Controlled benchmarks show that this reduced geometry, and in particular $R$, can be recovered from finite data even when the entries of the full high-dimensional operator are poorly estimated. Tests across data quantity, dimension, training horizon, spectral structure, and non-stationarity confirm that the relevant response geometry remains identifiable with far fewer samples than full-matrix recovery would require. Applied in moving windows to electrohysterogram, seizure EEG, freezing-of-gait, and unstable push-up inertial recordings, the method reveals systematic changes in the reduced non-normal geometry around known physiological or behavioral episodes. These changes involve shifts in $R$, changes in $\Delta$, or increased projection of the observed fluctuations onto the inferred response direction.
These applications show that the method can expose interpretable changes in local response geometry associated with known events, without treating the problem as one of supervised event detection.

 \end{abstract} 


\end{frontmatter}

% ============================================================
% ============================================================
\section{Introduction}
\label{sec:introduction}

A stable system can still erupt. A linear system whose eigenvalues all predict decay of disturbances can nonetheless amplify 
a perturbation by orders of magnitude before that decay
 sets in, provided its eigenvectors are sufficiently non-orthogonal. This non-normal transient growth is classical in hydrodynamic stability \cite{trefethen1993hydrodynamic,trefethen2005spectra,schmid2007nonmodal} and has since been recognized as a generic route to large finite-time responses in ecology \cite{neubert1997alternatives,neubert2002reactivity,tang2014reactivity}, neural circuits \cite{murphy2009balanced,hennequin2012nonnormal}, directed and hierarchical networks \cite{asllani2018nonnormal,asllani2018topological,nicoletti2018nonnormal,muolo2021synchronization}, non-Hermitian physics \cite{hatano1996localization}, and socio-economic influence networks that form bubbles or crashes below any conventional critical threshold \cite{sornette2003crash,sornette2023nonnormalbubbles}. The unifying point is that the size of a finite-time event can be set by geometry, not by eigenvalues, and the early-warning signatures usually attributed to proximity to a bifurcation \cite{scheffer2009earlywarnings,scheffer2012anticipating,lenton2008tipping}, including the still-debated critical slowing down before epileptic seizures \cite{maturana2020critical,wilkat2019noevidence}, can also arise from a stable system with non-normal local dynamics \cite{troude2024pseudobifurcations,troude2025unifying}.

The possible relevance of the non-normal mechanism shifts the empirical problem from estimating distance to an instability to identifying the local response geometry that can amplify perturbations while the system remains asymptotically stable. The difficulty is that this geometry is encoded in the local dynamical operator, which is rarely available in applications and must instead be inferred from multivariate observations. This is the typical situation in physiological and behavioral recordings, such as scalp electroencephalogram (EEG), inertial gait sensors, or multichannel electrohysterogram (EHG) signals \cite{goldberger2000physiobank,bloem2004fog,alexandersson2015ehg}. For such data, eigenvalues estimated from a fitted local model may characterize asymptotic decay, but they do not reveal the input and response directions through which finite-time amplification occurs. Variance-based reductions face the complementary limitation: they may identify energetic modes of variability without recovering the directed coupling geometry that generates them. Existing data-driven modelling tools infer governing equations, interaction matrices, or latent dynamics \cite{brunton2016sindy,rudy2017pdefind,course2023state,gao2024lagna}, but they do not directly target the low-dimensional non-normal response geometry itself.
The question we address is therefore precise: from a multivariate time series alone, can one infer the dominant two-dimensional input--response plane of a locally stable system, and measure how close this plane lies to the threshold for transient amplification, even when the full operator is not reliably identifiable entry by entry? We show that the answer is yes. The reduced geometry remains identifiable across data quantity, dimension, training horizon, spectral structure, and non-stationarity.

We therefore develop non-normal directional response inference, a data-driven framework for extracting transient-amplification geometry from multivariate time series. The method estimates local linear dynamics from sliding windows of the data, identifies the dominant two-dimensional plane through which perturbations are received and transiently amplified, and reduces the resulting geometry to interpretable diagnostics.
We compare three ways of identifying this plane: an eigenbasis-SVD baseline (M1), an optimization-based search for directed input--response coupling (M2), and a commutator-based construction targeting non-normality directly (M3). The dynamics restricted to the selected plane are then summarized by the reduced eigenvalue splitting $\Delta$, the reduced non-normality index $K$, and the scale-free ratio $R=K/K_c(\Delta)$, which locates the inferred geometry below, at, or above the two-dimensional threshold for transient amplification.

The paper proceeds in two stages. First, controlled synthetic benchmarks establish whether the proposed framework can recover the dominant non-normal plane and the normalized reduced non-normality ratio $R$ from finite data. We test robustness across fixed non-normality parameter scans, sample size, system dimension, training horizon, broad spectral structure, and time-varying non-normality. Second, we apply the same moving-window procedure to four empirical settings, ordered by decreasing evidential strength: uterine EHG activity, seizure EEG, freezing-of-gait acceleration, and push-up inertial recordings. The purpose of these applications is not to build universal event detectors, but to ask whether known physiological or behavioral episodes are accompanied by interpretable changes in local response geometry. The central claim is that non-normal amplification geometry can be inferred from time series alone and can explain finite-time responses in locally stable high-dimensional systems.

%%%%%%%%%%%%%%%%%%%%%%%%%%%%%%%%%%%%%%%%%%%%%%%%
\section{Motivation: stable systems can still amplify}
\label{sec:motivation}

Classical linear stability analysis describes the long-time fate of small perturbations. If all eigenvalues of a linearized continuous-time system lie in the stable half-plane, then perturbations eventually decay. This statement is correct, but it does not describe the full finite-time response. A stable non-normal system can first amplify selected perturbations before the eventual decay becomes visible.

A minimal example is the two-dimensional system
\begin{equation}
    \dot{\x} = A_\alpha \x,
    \qquad
    A_\alpha =
    \begin{pmatrix}
        -1 & \alpha \\
        0 & -2
    \end{pmatrix},
    \qquad
    \alpha > 0 .
    \label{eq:motivation_nonnormal_matrix}
\end{equation}
The eigenvalues of \(A_\alpha\) are \(-1\) and \(-2\), independently of \(\alpha\). Thus the system is asymptotically stable for every value of the coupling parameter \(\alpha\). The matrix is diagonalizable because the two eigenvalues are distinct, so the transient amplification discussed below does not rely on a defective matrix or on eigenvalue degeneracy. The matrix is nevertheless non-normal for \(\alpha\neq0\), since
\begin{equation}
    A_\alpha A_\alpha^\top \neq A_\alpha^\top A_\alpha .
    \label{eq:motivation_nonnormal_condition}
\end{equation}
The off-diagonal term \(\alpha\) creates a directional coupling between two stable modes.

\begin{figure*}[!htpb]
    \centering
    \includegraphics[width=\textwidth]{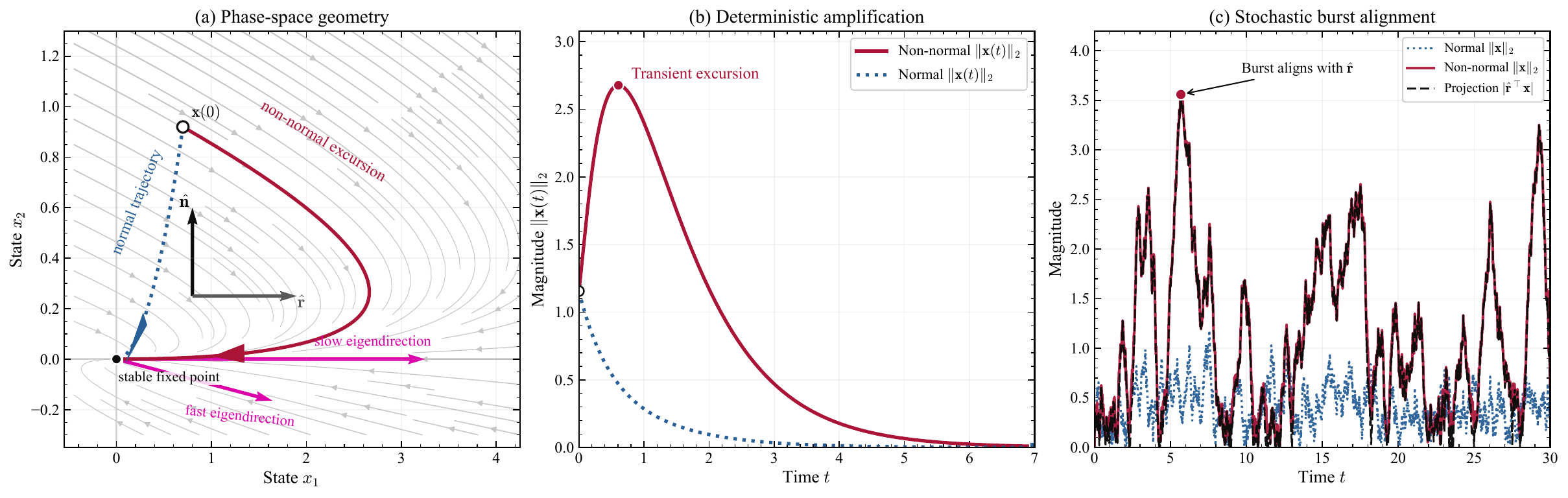}
    \caption{
    \textbf{Finite-time amplification in a stable non-normal system.}
    The non-normal system is \(\dot{\x}=A_{\rm nn}\x\), with
    \(A_{\rm nn}=\begin{pmatrix}-1&10\\0&-2\end{pmatrix}\).
    The normal comparison is \(\dot{\x}=A_{\rm normal}\x\), with
    \(A_{\rm normal}=\begin{pmatrix}-1&0\\0&-2\end{pmatrix}\).
    Both matrices have eigenvalues \(-1\) and \(-2\).
    \textbf{(a)} Phase-space trajectories from the initial condition
    \(\x(0)=(0.70,0.92)^\top\). The grey streamlines show the vector field of the non-normal system. The solid red curve is the non-normal trajectory and the dotted blue curve is the normal trajectory. The magenta arrows show the slow and fast eigendirections of \(A_{\rm nn}\). The directions \(\nhat\) and \(\rhat\) indicate the input and response directions used schematically in the reduced description.
    \textbf{(b)} Deterministic time series of the Euclidean norm \(\|\x(t)\|_2\) for the same two systems and the same initial condition as in panel~(a). The red marker denotes the maximum of the non-normal norm.
    \textbf{(c)} Stochastically forced trajectories generated from
    \(\dot{\x}=A\x+\sigma\bm{\xi}(t)\), with \(\sigma=0.55\), time step \(dt=0.01\), and the same noise realization for the normal and non-normal systems. The blue dotted curve is \(\|\x(t)\|_2\) for the normal system, the red curve is \(\|\x(t)\|_2\) for the non-normal system, and the black dashed curve is the projection \(|\rhat^\top\x(t)|\), where \(\rhat\) is the direction of the largest non-normal burst in the plotted stochastic trajectory.
    }
    \label{fig:motivation}
\end{figure*}

For the simple initial condition
\begin{equation}
    \x(0)=
    \begin{pmatrix}
        0 \\
        1
    \end{pmatrix},
    \label{eq:motivation_initial_condition}
\end{equation}
the solution is
\begin{equation}
       x_1(t)=\alpha\left(e^{-t}-e^{-2t}\right)   \qquad   x_2(t)=e^{-2t} ~.
    \label{eq:motivation_solution}
\end{equation}
The second component decays monotonically, but during its decay it feeds the first component.
The response \(x_1(t)\) first grows as $ x_1(t)=\alpha \left(t -{3 t^2 \over 2}
+ {\cal O}(t^3) \right)$ at short times, reaches its maximum at
\begin{equation}
    t_*=\log 2,
    \qquad
    x_1(t_*)=\frac{\alpha}{4},
    \label{eq:motivation_peak}
\end{equation}
and then decays to zero. Thus the transient response can be increased by increasing the non-normal coupling \(\alpha\), without moving either eigenvalue toward instability. The initial condition in \eqref{eq:motivation_initial_condition} is chosen to make the algebra transparent; the same qualitative behavior occurs for a generic starting point, as the phase portrait in \cref{fig:motivation} illustrates.

This example separates two notions that are often conflated. Spectral stability controls the long-time decay rate. Non-normal geometry controls the finite-time path taken by the perturbation before that decay is observed. In the notation used below, the spectral part of the reduced dynamics is summarized by the reduced eigenvalue splitting \(\Delta\), while the geometric part is summarized by the reduced non-normality index \(K\). These are the two fundamental contributors to finite-time amplification considered in this paper. Other observable effects sometimes discussed separately, such as one-step singular-vector gain, near-rank-one channel focusing, or cumulative finite-time response, are consequences of these two ingredients rather than independent amplification mechanisms. Their combination is reported through the normalized reduced non-normality ratio
\[
    R=K/K_c(\Delta),
\]
where \(K_c(\Delta)\) is the reduced threshold obtained from the two-dimensional real-eigenvalue calculation. The parameter \(R\) is therefore a reduced diagnostic: \(R<1\) means below the reduced threshold, \(R=1\) means at the reduced threshold, and \(R>1\) means above the reduced threshold.

The relevant directions are not necessarily eigenvectors. One direction acts as an input direction, denoted by \(\nhat\), while the induced response is expressed along a response direction, denoted by \(\rhat\). In the example above, perturbations initialized with a component in the second coordinate can generate a transient response in the first coordinate before both components decay. \ref{supp:two_dimensional_transient_growth} gives the corresponding discrete-time calculation for a canonical \(2\times2\) non-normal matrix. That calculation gives the threshold \(K_c(\Delta)\) used later to normalize the reduced non-normality index \(K\).

The same mechanism also appears under stochastic forcing. Consider
\begin{equation}
    \dot{\x} = A_\alpha \x + \sigma \bm{\xi}(t),
    \label{eq:motivation_stochastic}
\end{equation}
where \(\bm{\xi}(t)\) denotes temporally uncorrelated forcing and \(\sigma\) sets the forcing amplitude. Although the deterministic system is stable, repeated perturbations can be projected onto directions that undergo transient amplification. The resulting time series can therefore contain intermittent bursts whose dominant component is aligned with the response direction. A normal system with the same eigenvalues does not have the same finite-time input--response geometry.

\Cref{fig:motivation} illustrates the same distinction in phase space, deterministic time series, and stochastically forced time series. Panel~(a) shows that the non-normal trajectory can move away from the stable fixed point before returning to it, whereas the normal comparison decays directly. Panel~(b) shows the corresponding transient increase in \(\|\x(t)\|_2\) for the non-normal system. Panel~(c) shows that, under repeated forcing, the largest bursts in the non-normal system are concentrated along a response direction. These observations motivate the inference problem studied in this paper.

In empirical systems, the local operator is unknown, the state dimension can be high, and the amplification mechanism may be concentrated in a low-dimensional subspace. Eigenvalues alone do not identify the directions responsible for transient growth. Variance-based reductions also need not recover the directional coupling that produces amplification. The problem is therefore not only to estimate a stable operator from data, but also to infer the low-dimensional geometry through which stable dynamics can produce large finite-time or noise-driven responses.

The framework developed below follows this logic. From multivariate time series, we estimate a local linear operator, extract a two-dimensional non-normal response plane, and project the fitted dynamics onto that plane. The reduced operator is then used to compute the reduced eigenvalue splitting \(\Delta\), the reduced non-normality index \(K\), and the normalized reduced non-normality ratio \(R=K/K_c(\Delta)\). This provides a data-driven route from observed fluctuations to the reduced geometry that organizes them.
% ============================================================
\section{Data-driven inference of reduced non-normal geometry}
\label{sec:framework}

\subsection{The inference problem: recovering a reduced non-normal response geometry}

The preceding example shows that a stable linear system can exhibit large transient amplification when its eigenvectors are non-orthogonal. We now formulate the corresponding inference problem for data. Stated compactly, the problem is the following.
\par\smallskip
\noindent\textbf{Input.} A multivariate time series \(\{\x_k\}_{k=1}^{T}\), \(\x_k\in\R^N\), sampled from a locally stable high-dimensional system whose generating operator is unknown.
\par\smallskip
\noindent\textbf{Step 1 (fit).} Estimate a one-step linear operator $\widehat{F}$ from the data in the current window using ridge regression \eqref{eq:ridge_Fhat}.
\par\smallskip
\noindent\textbf{Step 2 (reduce).} Extract a two-dimensional orthonormal input--response plane \(Q=[\rhat,\nhat]\in\R^{N\times2}\), with input direction \(\nhat\) and response direction \(\rhat\), and project the fitted operator onto it, \(\Gam=Q^\top\Fhat Q\in\R^{2\times2}\).
\par\smallskip
\noindent\textbf{Output.} Three scalar diagnostics of \(\Gam\): the reduced eigenvalue splitting \(\Delta\), the reduced eigenvector non-orthogonality \(K\), and the normalized reduced non-normality ratio \(R=K/K_c(\Delta)\), where \(K_c(\Delta)\) is the two-dimensional transient-amplification threshold (\ref{supp:two_dimensional_transient_growth}). The values \(R<1\), \(R=1\), \(R>1\) place the inferred geometry below, at, or above that threshold.
\par\smallskip
The target of inference is this two-dimensional geometry, not entrywise recovery of the full \(N\times N\) operator. The synthetic benchmarks of \cref{sec:synthetic} establish that the geometry is recoverable even when the full operator is not.

\begin{figure*}[!htpb]
    \centering
    \includegraphics[width=\textwidth]{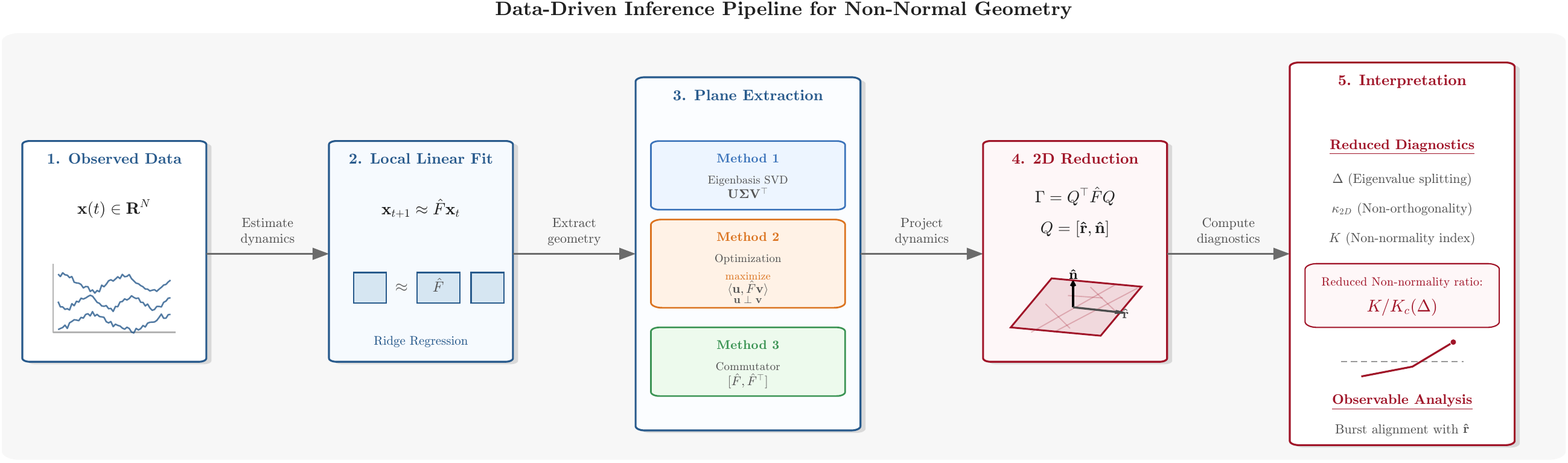}
    \caption{
    \textbf{Schematic of the inference workflow.}
    A multivariate time series \(\{\x_k\}\) is used to estimate a local one-step linear operator \(\Fhat\). A two-dimensional orthonormal basis \(Q=[\rhat,\nhat]\) is then extracted using one of the methods introduced in \cref{sec:methods}. The fitted operator is projected onto this plane to obtain the reduced matrix \(\Gam=Q^\top \Fhat Q\). The scalar diagnostics computed from \(\Gam\) are the reduced eigenvalue splitting \(\Delta\), the two-dimensional eigenvector-conditioning measure \(\kappa_{2D}\), the reduced non-normality index \(K\), and the normalized reduced non-normality ratio \(R=K/K_c(\Delta)\). Here \(K_c(\Delta)\) denotes the reduced threshold for transient amplification in the two-dimensional real-eigenvalue model.}
    \label{fig:pipeline_schematic}
\end{figure*}

\subsection{Observation model and local linear fit}
\label{sec:framework_observation_model}

Let \(\x(t)\in\R^N\) denote the observed state of a high-dimensional system. Over a finite observation window, we approximate the local dynamics by a stable linear system with additive stochastic forcing. In continuous time, this local model is written as
\begin{equation}
    \dot{\x}(t)
    =
    A\x(t)
    +
    \sqrt{2\delta}\,\bm{\xi}(t),
    \qquad
    \bm{\xi}(t)\sim\mathcal{N}(0,I_N),
    \label{eq:ct_observation_model}
\end{equation}
where \(A\in\R^{N\times N}\) is a stable local generator, \(I_N\) is the \(N\times N\) identity matrix, and \(\delta>0\) is the forcing intensity. For uniformly sampled data, and for the synthetic benchmarks used below, we use the discrete-time VAR(1) approximation
\begin{equation}
    \x_{k+1}
    =
    F\x_k+\bm{\eta}_k,
    \qquad
    \bm{\eta}_k\sim\mathcal{N}(0,\sigma^2 I_N),
    \qquad
    \rho(F)<1,
    \label{eq:dt_observation_model}
\end{equation}
where \(F\in\R^{N\times N}\) is the one-step linear map and \(\rho(F)\) denotes its spectral radius. When a continuous-time interpretation is appropriate and the sampling interval \(\Delta t\) is small, the two descriptions are formally related by
\begin{equation}
    F \approx e^{A\Delta t},
    \qquad
    A \approx \Delta t^{-1}\log(F).
    \label{eq:continuous_discrete_link}
\end{equation}

Given a uniformly sampled trajectory \(\{\x_k\}_{k=1}^{T}\), define
\begin{equation}
    X =
    \begin{bmatrix}
        \x_1 & \x_2 & \cdots & \x_{T-1}
    \end{bmatrix},
    \qquad
    Y =
    \begin{bmatrix}
        \x_2 & \x_3 & \cdots & \x_T
    \end{bmatrix}.
    \label{eq:XY_def}
\end{equation}
The fitted one-step map is obtained by ridge regression,
\begin{equation}
    \Fhat(\lambda)
    =
    YX^\top
    \left(XX^\top+\lambda I_N\right)^{-1},
    \qquad
    \lambda>0 .
    \label{eq:ridge_Fhat}
\end{equation}

For multiple independent trajectories generated under the same operator, the corresponding \(X\) and \(Y\) matrices are concatenated columnwise before applying
\eqref{eq:ridge_Fhat}. In the synthetic benchmarks, \(M\) denotes the number
of such independent training trajectories, and \(T_{\mathrm{train}}\) denotes
the number of time samples in each trajectory. Thus, \(M>1\) is used only in
controlled simulations where repeated trajectories from the same known operator
can be generated. In empirical recordings, one usually observes a single long
time series rather than repeated independent realizations of the same local
dynamics. We therefore set \(M=1\) for each empirical window and estimate a
separate window-dependent operator from the consecutive sample pairs inside
that window. When several comparable events are available, such as multiple
seizures, freezing episodes, contractions, or repeated movement bursts, they are
used to assess repeatability of the diagnostics across events or recordings.
They are not treated as independent trajectories for a single fitted operator
unless an explicit pooling assumption is stated.

The regularization parameter \(\lambda\) controls the finite-sample bias--variance tradeoff. In synthetic benchmarks, it is selected using predictive accuracy and stability of the fitted map. In empirical moving-window analyses, additional mild stabilization or screening is applied when the local regression problem is poorly conditioned. These empirical safeguards are described in \cref{sec:empirical}, because they depend on the dataset and are not part of the mathematical definition of the reduced diagnostics.

\subsection{Target of inference}
\label{sec:framework_target}

The goal is not to recover every entry of the full operator with equal accuracy. Instead, the goal is to identify the low-dimensional geometry that organizes the dominant transient response. We therefore seek a two-dimensional orthonormal basis
\begin{equation}
    Q =
    \begin{bmatrix}
        \rhat & \nhat
    \end{bmatrix}
    \in \R^{N\times 2},
    \qquad
    Q^\top Q=I_2 .
    \label{eq:Q_def}
\end{equation}
Here \(\nhat\) denotes the inferred input direction in which perturbations or fluctuations are most susceptible to transient amplification, while \(\rhat\) denotes the associated response direction. The interpretation is that a component along \(\nhat\) is preferentially mapped by the fitted dynamics into the direction \(\rhat\).

Once \(Q\) has been identified, the fitted operator is projected onto this plane:
\begin{equation}
    \Gam
    =
    Q^\top \Fhat Q
    \in \R^{2\times 2}.
    \label{eq:reduced_operator}
\end{equation}
All reduced diagnostics are computed from the same \(2\times2\) matrix \(\Gam\), regardless of how the plane \(Q\) was extracted. This common projection step makes it possible to compare different plane-extraction methods using the same reduced operator.

\subsection{Reduced diagnostics}
\label{sec:framework_diagnostics}

Let \(\lambda_1,\lambda_2\) denote the eigenvalues of the reduced operator \(\Gam\). We first define the normalized reduced eigenvalue splitting
\begin{equation}
    \Delta
    =
    \left|
    \frac{\lambda_1-\lambda_2}{\lambda_1+\lambda_2}
    \right|.
    \label{eq:Delta_def}
\end{equation}
The quantity \(\Delta\) measures the separation of the two reduced eigenvalues relative to their mean scale. Small \(\Delta\) corresponds to a nearly degenerate reduced spectrum, whereas larger values indicate more separated reduced eigenvalues.

Let \(\bm p_1,\bm p_2\) be normalized right eigenvectors of \(\Gam\). Their non-orthogonality is measured by
\begin{equation}
    \kappa_{2D}
    =
    \sqrt{
    \frac{1+\left|\ip{\bm p_1}{\bm p_2}\right|}
         {1-\left|\ip{\bm p_1}{\bm p_2}\right|}
    } .
    \label{eq:kappa2D_def}
\end{equation}
For an orthogonal reduced eigenbasis, \(\kappa_{2D}=1\). As the two reduced eigenvectors become nearly parallel, \(\kappa_{2D}\) increases.

We then define the reduced non-normality index
\begin{equation}
    K
    =
    \frac{\kappa_{2D}-\kappa_{2D}^{-1}}{2}.
    \label{eq:K_def}
\end{equation}
This scalar is zero for an orthogonal reduced eigenbasis and increases with the non-orthogonality of the two reduced eigenvectors. It therefore measures the geometric component of transient amplification inside the inferred two-dimensional plane.

The reduced threshold \(K_c(\Delta)\) is obtained from the two-dimensional real-eigenvalue calculation summarized in \ref{supp:two_dimensional_transient_growth}. In this setting,
\begin{equation}
    K_c(\Delta)
    =
    \left[
    \frac{\sqrt{1-\Delta^2}}
         {1-\sqrt{1-\Delta^2}}
    \right]^{1/2},
    \qquad
    0\leq \Delta < 1 .
    \label{eq:Kc_def}
\end{equation}
We compare the measured reduced non-normality \(K\) with this threshold through the normalized reduced non-normality ratio
\begin{equation}
    R
    =
    \frac{K}{K_c(\Delta)} .
    \label{eq:KKc_ratio_def}
\end{equation}
The parameter \(R\) should be interpreted as a reduced diagnostic. Values \(R<1\) indicate that the reduced geometry lies below the two-dimensional transient-amplification threshold defined by \eqref{eq:Kc_def}. Values \(R=1\) correspond to that reduced threshold, and values \(R>1\) indicate that the reduced geometry lies above it. Thus \(R\) controls whether the inferred two-dimensional dynamics are below or above the reduced threshold; \(R\) itself is not the phenomenon of criticality. The threshold in \eqref{eq:Kc_def} is specific to the two-dimensional real-eigenvalue reduction and should not be interpreted as a universal stability boundary for the full \(N\)-dimensional system.

\paragraph{Conditioning of the operator versus conditioning of eigenvectors.}
We distinguish four related but non-equivalent quantities. First, the matrix condition number
\begin{equation}
    \kappa(\Fhat)
    =
    \frac{\sigma_{\max}(\Fhat)}
         {\sigma_{\min}(\Fhat)}
    \label{eq:matrix_condition_number}
\end{equation}
measures the spread of singular values of the fitted operator. It is sensitive to anisotropic gain and near-singularity, but it is not by itself a measure of eigenvector non-orthogonality. Second, if
\begin{equation}
    \Fhat P=P\Lambda ,
    \label{eq:eigenbasis_conditioning_relation}
\end{equation}
then
\begin{equation}
    \kappa(P)=\norm{P}_2\norm{P^{-1}}_2
    \label{eq:eigenbasis_condition_number}
\end{equation}
measures the conditioning of the eigenvector matrix \(P\). This quantity is tied to the non-orthogonality of the eigenvectors of the full fitted operator \(\Fhat\). Third, \(\kappa_{2D}\) in \eqref{eq:kappa2D_def} measures the non-orthogonality of the eigenvectors of the reduced operator \(\Gam=Q^\top\Fhat Q\). Fourth, \(K\) in \eqref{eq:K_def} is a scalar reparameterization of \(\kappa_{2D}\) used to compare the reduced geometry with the threshold \(K_c(\Delta)\). Thus \(\kappa(\Fhat)\), \(\kappa(P)\), \(\kappa_{2D}\), and \(K\) are different objects and should not be interchanged.

\subsection{Validation metrics in synthetic data}
\label{sec:framework_validation}

When a known reference operator \(F\) is available, the inference can be evaluated at three levels. First, full-operator recovery is measured by the relative Frobenius error
\begin{equation}
    \mathrm{relErr}(\Fhat)
    =
    \frac{\norm{\Fhat-F}_F}{\norm{F}_F}.
    \label{eq:relerr_def}
\end{equation}
This diagnostic quantifies the accuracy of the high-dimensional regression problem.

Second, plane recovery is measured using the largest principal angle between the inferred plane \(Q\) and a reference plane \(Q_{\mathrm{ref}}\). We define
\begin{equation}
    \theta(Q,Q_{\mathrm{ref}})
    =
    \arccos
    \left[
    \sigma_{\min}
    \left(
    Q^\top Q_{\mathrm{ref}}
    \right)
    \right],
    \label{eq:principal_angle_def}
\end{equation}
where \(\sigma_{\min}\) denotes the smallest singular value. The matrix \(Q^\top Q_{\mathrm{ref}}\) contains the inner products between the two orthonormal bases. Its singular values are the cosines of the principal angles between the two planes, so \eqref{eq:principal_angle_def} reports the largest angular mismatch.

Third, mechanism-level recovery is assessed using the reduced diagnostics and observable correlations. For a scalar observable such as the mean field
\begin{equation}
    m_k
    =
    \frac{1}{N}\bm{1}^\top \x_k,
    \label{eq:meanfield_def}
\end{equation}
we use a plane-invariant correlation score
\begin{equation}
    c_{\mathrm{plane}}
    =
    \max_{\norm{\bm a}=1}
    \left|
    \corr
    \left(
        m_k,\,
        \bm a^\top Q^\top \x_k
    \right)
    \right|.
    \label{eq:cplane_def}
\end{equation}
This score asks whether the inferred two-dimensional plane contains a direction whose activity is strongly related to the observable. Because the maximization is over all unit directions inside the plane, the score depends on the plane itself and not on an arbitrary choice of basis within the plane.

These three levels answer different questions. The operator error in \eqref{eq:relerr_def} measures recovery of the full \(N\times N\) map. The principal angle in \eqref{eq:principal_angle_def} measures recovery of the reduced response plane. The reduced diagnostics and \(c_{\mathrm{plane}}\) measure whether that plane captures an amplification-related component visible in the data. This distinction is important because the reduced response plane can be recovered well even when the full operator is not recovered entry by entry.

\subsection{Use in empirical data}
\label{sec:framework_empirical_use}

For empirical datasets, the reference operator and reference response plane are unavailable. We therefore do not report \(\mathrm{relErr}(\Fhat)\) or reference principal-angle errors. Instead, the evidence is based on internal consistency and alignment with independently observed events: agreement between independent extraction methods, persistence of the reduced diagnostics across neighboring windows, alignment with event times or activity envelopes, and exclusion of degenerate reduced matrices.

In empirical data, \(R=K/K_c(\Delta)\) should be read as a local reduced diagnostic of amplification-prone geometry. A rise in \(R\) does not by itself prove causality or establish a universal event detector. Its interpretation depends on whether the rise is reproducible across neighboring windows, whether the optimization-based extraction (M2) and commutator-based extraction (M3) agree, whether the reduced eigenvalue splitting \(\Delta\) remains non-degenerate, and whether the inferred response direction is related to an observed signal. This conservative interpretation is used throughout the empirical demonstrations in \cref{sec:empirical}.

% ============================================================
\section{Plane-extraction methods}
\label{sec:methods}

\subsection{Three methods for extracting the dominant two-dimensional response plane}

The reduced diagnostics introduced in \cref{sec:framework_diagnostics} require a two-dimensional plane \(Q=[\rhat,\nhat]\) on which the fitted operator \(\Fhat\) is projected. We compare three ways of extracting this plane from the same fitted operator. The first method, \emph{eigenbasis-SVD baseline extraction (M1)}, applies a singular-value decomposition (SVD) to the matrix of right eigenvectors of $\widehat{F}$ and uses the resulting dominant directions to define the reduced plane. The second method, \emph{optimization-based directed-coupling extraction (M2)}, searches directly for an orthonormal pair of directions with large transverse action under \(\Fhat\). The third method, \emph{commutator-based plane extraction (M3)}, identifies a two-dimensional plane from the symmetric commutator of \(\Fhat\).

For compactness, we refer to these methods as M1, M2, and M3 in figures and tables. Their full names are
\[
\begin{aligned}
\mathrm{M1} &:\quad \text{eigenbasis-SVD baseline extraction},\\
\mathrm{M2} &:\quad \text{optimization-based directed-coupling extraction},\\
\mathrm{M3} &:\quad \text{commutator-based plane extraction}.
\end{aligned}
\]
Each method takes \(\Fhat\) as input and returns either an ordered orthonormal pair
\[
    Q=[\rhat,\nhat]\in\R^{N\times 2},
    \qquad
    Q^\top Q=I_2,
\]
or, in the case of M3, a two-dimensional plane that is subsequently ordered into the pair \((\rhat,\nhat)\) when an input--response interpretation is needed. In all cases, the reported reduced operator is
\[
    \Gam = Q^\top \Fhat Q,
\]
and the diagnostics \(\Delta\), \(\kappa_{2D}\), \(K\), and \(R=K/K_c(\Delta)\) are computed from \(\Gam\) as defined in \cref{sec:framework_diagnostics}. Thus the three methods differ only in how the plane \(Q\) is extracted; the final projection and diagnostic definitions are the same.

% ------------------------------------------------------------
\subsection{Method 1: eigenbasis-SVD baseline extraction (M1)}
\label{sec:method1}

The eigenbasis-SVD baseline extraction (M1) provides a reference construction based on the eigenspace geometry of the fitted operator. Starting from \(\Fhat\), we compute a numerical eigendecomposition
\begin{equation}
    \Fhat P = P\Lambda,
    \label{eq:m1_eigendecomposition}
\end{equation}
where the columns of \(P\) are right eigenvectors and \(\Lambda\) is the corresponding diagonal matrix of eigenvalues.

The conditioning of the eigenvector matrix is obtained from the singular value decomposition   
\begin{equation}
    P = U\Sigma V^\top,
    \qquad
    \Sigma=\mathrm{diag}(\sigma_1,\ldots,\sigma_N),
    \qquad
    \sigma_1\geq \cdots \geq \sigma_N >0 .
    \label{eq:m1_svd}
\end{equation}
We define
\begin{equation}
     \kappa(P)
    =
    \frac{\sigma_1}{\sigma_N},
    \label{eq:m1_kappaeig}
\end{equation}
which is equivalent to (\ref{eq:eigenbasis_condition_number}).
This quantity measures the conditioning of the eigenvector matrix \(P\). Large $\kappa(P)$ indicates that the eigenvectors of \(\Fhat\) are close to being linearly dependent. It is therefore a measure of eigenvector geometry. It is not the matrix condition number
\[
    \kappa(\Fhat)=
    \frac{\sigma_{\max}(\Fhat)}
         {\sigma_{\min}(\Fhat)},
\]
which measures the singular-value spread of the fitted operator itself.

In M1, the non-normal direction is taken to be the left singular vector of \(P\) associated with its smallest singular value:
\begin{equation}
    \nhat = U_{:,N}.
    \label{eq:m1_nhat}
\end{equation}
The corresponding response direction is obtained by applying \(\Fhat\) to \(\nhat\) and removing the component parallel to \(\nhat\):
\begin{equation}
    \widetilde{\bm r}
    =
    \left(I_N-\nhat\nhat^\top\right)\Fhat\nhat,
    \qquad
    \rhat
    =
    \frac{\widetilde{\bm r}}
         {\norm{\widetilde{\bm r}}}.
    \label{eq:m1_rhat}
\end{equation}
The extracted basis and the reduced operator are
\begin{equation}
    Q_{\mathrm{M1}}
    =
    [\rhat,\nhat],
    \qquad
    \Gam_{\mathrm{M1}}
    =
    Q_{\mathrm{M1}}^\top \Fhat Q_{\mathrm{M1}} .
    \label{eq:m1_gamma}
\end{equation}

M1 is useful because it connects directly to the classical eigenvector picture of non-normality. Its limitation is that it relies on the eigendecomposition of \(\Fhat\). When the fitted operator is noisy, nearly defective, or strongly non-normal, small perturbations in \(\Fhat\) can produce large changes in its eigenvectors. For this reason, M1 is used as a baseline rather than as the main estimator of the response plane.

% ------------------------------------------------------------
\subsection{Method 2: optimization-based directed-coupling extraction (M2)}
\label{sec:method2}

The optimization-based directed-coupling extraction (M2) estimates the input and response directions directly from the fitted operator, without using an eigendecomposition. The method searches for an orthonormal pair of directions for which the action of \(\Fhat\) maps one direction strongly into the other.

We define \((\rhat,\nhat)\) as a solution of
\begin{equation}
    (\rhat,\nhat)
    =
    \arg\max_{\substack{\norm{\bm u}=1,\ \norm{\bm v}=1\\
    \ip{\bm u}{\bm v}=0}}
    \ip{\bm u}{\Fhat \bm v}.
    \label{eq:m2_objective}
\end{equation}
Here \(\nhat\) is the input direction and \(\rhat\) is the response direction. The objective in \eqref{eq:m2_objective} measures the component of \(\Fhat\nhat\) transverse to \(\nhat\) and aligned with \(\rhat\).

A practical solver is alternating maximization with orthogonal projection. Given a current vector \(\bm v\), the maximizing update for \(\bm u\) is
\begin{equation}
    \bm u
    \leftarrow
    \frac{
    (I_N-\bm v\bm v^\top)\Fhat\bm v
    }{
    \norm{(I_N-\bm v\bm v^\top)\Fhat\bm v}
    } .
    \label{eq:m2_update_u}
\end{equation}
Given a current vector \(\bm u\), the corresponding update for \(\bm v\) is
\begin{equation}
    \bm v
    \leftarrow
    \frac{
    (I_N-\bm u\bm u^\top)\Fhat^\top\bm u
    }{
    \norm{(I_N-\bm u\bm u^\top)\Fhat^\top\bm u}
    } .
    \label{eq:m2_update_v}
\end{equation}
The updates are iterated until the objective in \eqref{eq:m2_objective}, or equivalently the extracted plane, changes by less than a prescribed numerical tolerance. Multiple initializations are used to reduce sensitivity to local optima.

After convergence, we impose a consistent response-direction convention. Once \(\nhat\) has been identified, we set
\begin{equation}
    \widetilde{\bm r}
    =
    (I_N-\nhat\nhat^\top)\Fhat\nhat,
    \qquad
    \rhat
    =
    \frac{\widetilde{\bm r}}
         {\norm{\widetilde{\bm r}}}.
    \label{eq:m2_rhat}
\end{equation}
The basis and reduced operator are then
\begin{equation}
    Q_{\mathrm{M2}}
    =
    [\rhat,\nhat],
    \qquad
    \Gam_{\mathrm{M2}}
    =
    Q_{\mathrm{M2}}^\top \Fhat Q_{\mathrm{M2}} .
    \label{eq:m2_gamma}
\end{equation}

M2 targets the directed amplification geometry directly. It does not require the eigenvectors of \(\Fhat\), and therefore avoids one source of numerical instability present in M1. It also returns an ordered pair \((\rhat,\nhat)\), which is useful in empirical data because the response direction can be compared with observable activity, event-aligned signals, or collective coordinates.

% ------------------------------------------------------------
\subsection{Method 3: commutator-based plane extraction (M3)}
\label{sec:method3}

The commutator-based plane extraction (M3) identifies a two-dimensional non-normal plane from the symmetric commutator of the fitted operator. We define
\begin{equation}
    B
    =
    \Fhat\Fhat^\top
    -
    \Fhat^\top\Fhat .
    \label{eq:m3_commutator}
\end{equation}
For a normal operator, \(\Fhat\Fhat^\top=\Fhat^\top\Fhat\), so \(B=0\). Thus, \(B\) measures the departure of \(\Fhat\) from normality in the sense of non-commutation with its transpose. Since \(B\) is symmetric, its leading eigenspaces can be computed through a symmetric eigenproblem.

Let \(\bm w_+\) denote the eigenvector associated with the largest positive eigenvalue of \(B\), and let \(\bm w_-\) denote the eigenvector associated with the most negative eigenvalue of \(B\). The commutator plane is
\begin{equation}
    Q_{\mathrm{comm}}
    =
    [\bm w_+,\bm w_-].
    \label{eq:m3_comm_plane}
\end{equation}
If necessary, the two columns are orthonormalized numerically before projection. Projecting the fitted operator onto this plane gives
\begin{equation}
    \Gam_{\mathrm{comm}}
    =
    Q_{\mathrm{comm}}^\top \Fhat Q_{\mathrm{comm}} .
    \label{eq:m3_comm_gamma}
\end{equation}

Unlike M2, the commutator construction returns a plane rather than a uniquely ordered input--response pair. When only scalar reduced diagnostics are needed, they can be computed from \(\Gam_{\mathrm{comm}}\). When an ordered pair is needed, for example to compare the response direction with an observable signal, we solve the two-dimensional version of the directed-coupling problem inside the commutator plane.

Concretely, let \((\bm r_c,\bm n_c)\in\R^2\times\R^2\) solve
\begin{equation}
    (\bm r_c,\bm n_c)
    =
    \arg\max_{\substack{\norm{\bm u}=1,\ \norm{\bm v}=1\\
    \ip{\bm u}{\bm v}=0}}
    \ip{\bm u}{\Gam_{\mathrm{comm}}\bm v}.
    \label{eq:m3_internal_ordering}
\end{equation}
The corresponding directions in the original state space are
\begin{equation}
    \rhat = Q_{\mathrm{comm}}\bm r_c,
    \qquad
    \nhat = Q_{\mathrm{comm}}\bm n_c,
    \qquad
    Q_{\mathrm{M3}}=[\rhat,\nhat].
    \label{eq:m3_lifted_directions}
\end{equation}
For consistency with M1 and M2, the reduced operator used for the reported M3 diagnostics is
\begin{equation}
    \Gam_{\mathrm{M3}}
    =
    Q_{\mathrm{M3}}^\top \Fhat Q_{\mathrm{M3}} .
    \label{eq:m3_gamma}
\end{equation}
This is the same commutator-plane projection expressed in the internally ordered basis.

The extremal eigenvalues of \(B\) provide an additional marker of non-normal structure. In the synthetic benchmarks, we report the leading positive and negative commutator eigenvalues separately from the reduced diagnostics. These eigenvalues of \(B\) should not be confused with the ridge parameter \(\lambda\) used to estimate \(\Fhat\), with the eigenvalues of the fitted operator \(\Fhat\), or with the eigenvalues of the reduced matrix \(\Gam\).

M3 is useful because the plane-extraction step is based on a symmetric matrix, even though the fitted operator \(\Fhat\) is generally non-normal. Its limitation is that, when the fitted dynamics are only weakly non-normal, the commutator signal can be small and the dominant plane can be less identifiable. In the moderately and strongly non-normal regimes considered below, M3 provides an independent check on the plane obtained by M2.

% ============================================================
% ============================================================

\section{Synthetic validation} 
\label{sec:synthetic}  

\subsection{Synthetic benchmark protocol and validation ladder}

We first test the reduced non-normality diagnostics on controlled synthetic data. In this setting, the full operator is known, so the reference response plane, the reduced operator, and the ratio \(R=K/K_c(\Delta)\) can be computed directly. The construction also allows the strength of the non-normal geometry to be varied while keeping the eigenvalue spectrum, forcing amplitude, and orientation template fixed. This separates changes caused by the reduced non-normal geometry from changes caused by the stochastic forcing or by unrelated changes in the spectrum.  The central parameter reported in this section is the normalized reduced non-normality ratio \(R=K/K_c(\Delta)\) of \eqref{eq:KKc_ratio_def}, where \(K\) is the reduced non-normality index and \(K_c(\Delta)\) the threshold set by the reduced eigenvalue splitting \(\Delta\) through \eqref{eq:Kc_def}.  The scalar \(\kappa\) used below is only an internal construction parameter used to generate a family of increasingly anisotropic operators. It should not be confused with \(\kappa(\Fhat)\), with the eigenvector-basis condition number \(\kappa(P)\), or with the reduced eigenvector-conditioning measure \(\kappa_{2D}\). We therefore present the synthetic transition in terms of \(R\), rather than in terms of the internal parameter \(\kappa\). We organise these tests as a ladder, each rung establishing one property of the inferred diagnostic: 
\begin{enumerate}
\item[(i)] the transition controlled by \(R\) exists in the reference reduction (\cref{sec:synthetic_var_benchmark}); 
\item[(ii)] the plane and \(R\) are recoverable once \(F\) is unknown and must be estimated from finite data (\cref{sec:synthetic_method_comparison}); 
\item[(iii)] recovery scales favourably with the amount of data and the ambient dimension, well below the \(N^2\) samples needed to identify the full operator (\cref{sec:synthetic_scaling}); 
\item[(iv)] the diagnostic can be tracked through time in a causal rolling window (\cref{sec:synthetic_timevarying_revised}); and 
\item[(v)] it stays identifiable when the operator carries a broad singular-value spectrum rather than a single small mode (\cref{sec:synthetic_broad_singular_values}). 
\end{enumerate}
The empirical demonstrations in \cref{sec:empirical} then apply exactly the moving-window construction validated by the last two rungs.

\subsection{VAR(1) benchmark} \label{sec:synthetic_var_benchmark}  

Synthetic trajectories are generated from the stable discrete-time VAR(1) model 
\begin{equation}     \x_{k+1}     =     F(\kappa)\x_k+\bm{\eta}_k,     \qquad     \bm{\eta}_k\sim\mathcal{N}(0,\sigma^2 I_N),   
  \qquad     \rho(F(\kappa))<1 .    
   \label{eq:synthetic_var1} 
   \end{equation} 
  Here \(F(\kappa)\in\mathbb{R}^{N\times N}\) is a stable one-step linear map and \(\kappa\) is the internal anisotropy parameter. We construct 
  \begin{equation}     
  F(\kappa)     =     U\,\Sigma(\kappa)\,V^\top     \Lambda     V\,\Sigma(\kappa)^{-1}U^\top ,     
  \label{eq:synthetic_Fkappa} 
  \end{equation} 
  where \(U\) and \(V\) are fixed orthogonal matrices, \(\Lambda\) is a fixed diagonal matrix of stable eigenvalues, and \(\Sigma(\kappa)\) is a diagonal anisotropy matrix. Within the parameter scan, \(U\), \(V\), \(\Lambda\), and the baseline singular-value scales are fixed. Only \(\kappa\) is varied. The spectral radius is fixed at \(\rho=0.95\), and the remaining eigenvalues are sampled inside the stable disk. Thus the observed changes across the parameter scan are driven by the changing non-normal geometry, not by movement of the spectrum toward instability.  For each trajectory, we examine three scalar projections. The first is the empirical mean field \(m_k=\tfrac{1}{N}\bm{1}^\top\x_k\) of \eqref{eq:meanfield_def}. The second is the input coordinate \(\nhat^\top\x_k\), and the third is the response coordinate \(\rhat^\top\x_k\). In this benchmark, the directions \((\rhat,\nhat)\) are extracted from the known operator \(F(\kappa)\). This isolates the effect of the reduced non-normal geometry from the additional uncertainty introduced by estimating the operator from finite data.  

\begin{figure*}[!htpb]    
 \centering     \includegraphics[width=\textwidth]{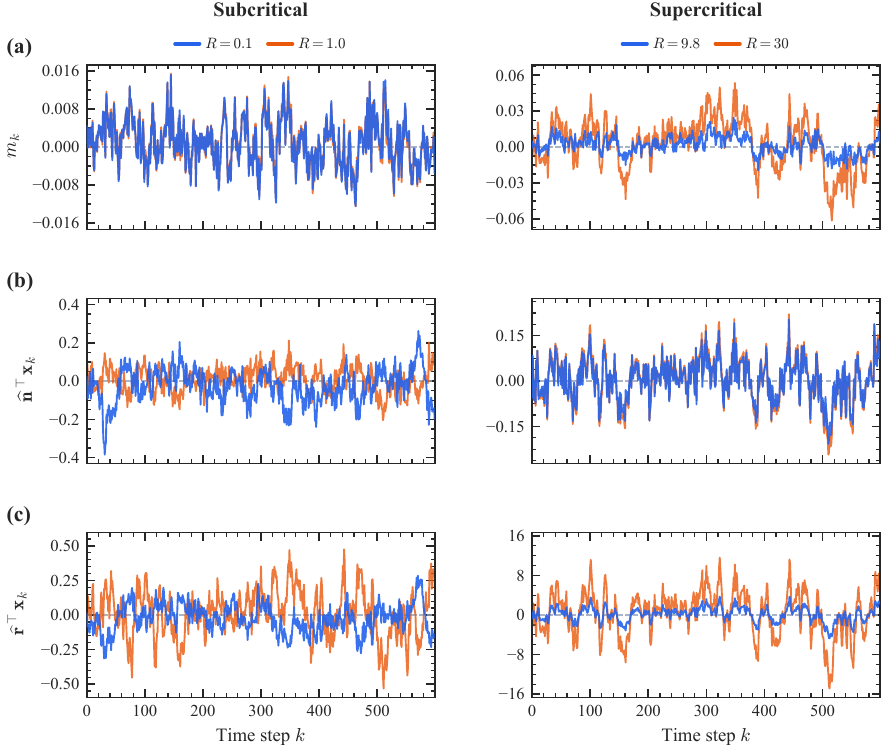}     
 \caption{     \textbf{Synthetic VAR(1) trajectories across values of the ratio \(R\).}     Trajectories are generated from \(\x_{k+1}=F(\kappa)\x_k+\bm{\eta}_k\), with \(N=200\), spectral radius \(\rho(F)=0.95\), \(\bm{\eta}_k\sim\mathcal{N}(0,\sigma^2 I_N)\), \(\sigma=0.05\), and \(T=600\) time steps. The same noise realization is used for the four representative cases. The operator \(F(\kappa)\) is constructed according to (\ref{eq:synthetic_Fkappa})
 with fixed orthogonal matrices \(U,V\), a fixed stable diagonal spectrum \(\Lambda\), and a varied diagonal anisotropy matrix \(\Sigma(\kappa)\). The plots correspond to \(R=0.1\), \(R=1\), \(R=9.8\), and \(R=30\), where \(R=K/K_c(\Delta)\) and $K_c$ is given by expression (\ref{eq:Kc_def}). The left column shows the two cases \(R=0.1\) and \(R=1\), and the right column shows the two cases \(R=9.8\) and \(R=30\) above the reduced threshold $R=1$. \textbf{(a)} Mean field \(m_k=N^{-1}\bm{1}^\top\x_k\). \textbf{(b)} Input coordinate \(\nhat^\top\x_k\). \textbf{(c)} Response coordinate \(\rhat^\top\x_k\). The directions \(\nhat\) and \(\rhat\) are extracted from the known operator \(F(\kappa)\) for each value of \(\kappa\).     }     
 \label{fig:synthetic_transition} 
 \end{figure*} 
 
\Cref{fig:synthetic_transition} shows representative trajectories for different values of $R$ defined in \eqref{eq:KKc_ratio_def}. For $R=0.1$ and $R=1$, the mean field and the two reduced coordinates remain small under the chosen forcing amplitude. For $R=9.8$ and $R=30$, the response coordinate $\rhat^\top\x_k$ 
 develops much larger excursions than the input coordinate $\nhat^\top\x_k$. Thus, as the reduced geometry crosses the transient-amplification threshold, fluctuations entering along the input direction are increasingly expressed along a nearly transverse response direction before eventually decaying.
 
 \begin{figure}[!htpb]     
 \centering     \includegraphics[width=\columnwidth]{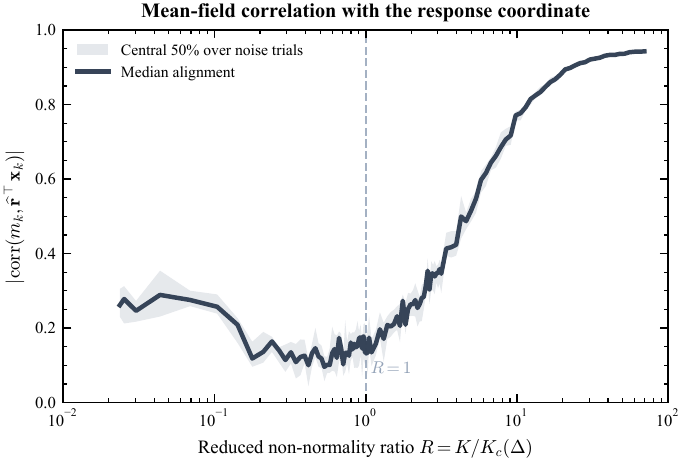}    
  \caption{     \textbf{Mean-field correlation with the response coordinate across the non-normal transition.}     The horizontal axis is the normalized reduced non-normality ratio \(R=K/K_c(\Delta)\), computed from the known reduced operator \(Q^\top F(\kappa)Q\). The vertical axis is the absolute correlation \(|\corr(m_k,\rhat^\top\x_k)|\) between the mean field \(m_k=N^{-1}\bm{1}^\top\x_k\) and the response coordinate \(\rhat^\top\x_k\). The parameter scan uses the same synthetic VAR(1) construction as in \cref{fig:synthetic_transition}, with \(N=200\), \(\rho(F)=0.95\), \(\sigma=0.05\), and \(T=1000\) time steps per realization. The internal anisotropy parameter \(\kappa\) is sampled at 116 values, with denser sampling below and around \(R=1\). For each value of \(\kappa\), the plotted line is the median over 16 independent noise realizations and the shaded band is the interquartile interval. The vertical dashed line marks \(R=1\).     }     
  \label{fig:observable_alignment} 
  \end{figure} 
  
  \Cref{fig:observable_alignment} summarizes the same transition over the parameter scan. The alignment 
  \begin{equation}     
  \left|     \corr\!\left(     m_k,\,     \rhat^\top\x_k     \right)     \right|     \label{eq:synthetic_reaction_corr}
  \end{equation} 
  is small for \(R<1\) and increases as \(R\) grows above $1$. This shows that the response direction becomes progressively more visible in the observed mean field as the non-normal geometry strengthens. The relevant control variable is $R=K/K_c(\Delta)$, which expresses the strength of the reduced non-normal geometry directly, rather than the internal anisotropy parameter $\kappa$ used only to generate the synthetic operators. Figures \ref{fig:synthetic_transition} and \ref{fig:observable_alignment} first validate the controlled synthetic construction. Because $F(\kappa)$ and the reference response plane are known, these plots do not yet test finite-data operator inference. They instead show that the constructed family realizes the intended transition: increasing $R$ turns a weakly expressed response coordinate into a strongly amplified one. The next subsection removes access to the true operator and asks whether the reduced geometry can still be recovered from observed trajectories alone.

% ------------------------------------------------------------
% ------------------------------------------------------------
\subsection{Recovery from estimated operators}
\label{sec:synthetic_method_comparison}

We next test the full inference problem. In this setting, the operator \(F\) is treated as unknown and must be estimated from observed trajectories. For each value of the internal anisotropy parameter \(\kappa\), training trajectories are generated from \eqref{eq:synthetic_var1}, a fitted operator \(\Fhat\) is obtained by ridge regression using \eqref{eq:ridge_Fhat}, and the three plane-extraction methods introduced in \cref{sec:methods} are applied to \(\Fhat\). The reference plane is obtained by applying optimization-based directed-coupling extraction (M2) directly to the known operator \(F\). To compare with the empirical setting, we show both the \(M=25\) ensemble benchmark and the \(M=1\) single-trajectory case.

The reference ratio is denoted by
\begin{equation}
    R_{\mathrm{true}}
    =
    \frac{K_{\mathrm{true}}}
         {K_c(\Delta_{\mathrm{true}})} ,
    \label{eq:synthetic_Rtrue}
\end{equation}
where \(K_{\mathrm{true}}\) and \(\Delta_{\mathrm{true}}\) are computed from the reference reduced operator
\[
    \Gam_{\mathrm{true}}
    =
    Q_{\mathrm{true}}^\top F Q_{\mathrm{true}} .
\]
For an estimated operator \(\Fhat\) and an extracted plane \(Q\), the corresponding inferred value is denoted by \(\widehat R\). The inferred plane is compared with the reference plane \(Q_{\mathrm{true}}\) using the largest principal angle \(\theta(Q,Q_{\mathrm{true}})\) of \eqref{eq:principal_angle_def}, with \(Q_{\mathrm{ref}}=Q_{\mathrm{true}}\), and the observable content of the recovered plane is measured using the plane-invariant alignment score \(c_{\mathrm{plane}}\) defined in \eqref{eq:cplane_def}.

\begin{figure*}[!htpb]
    \centering
    \includegraphics[width=\textwidth]{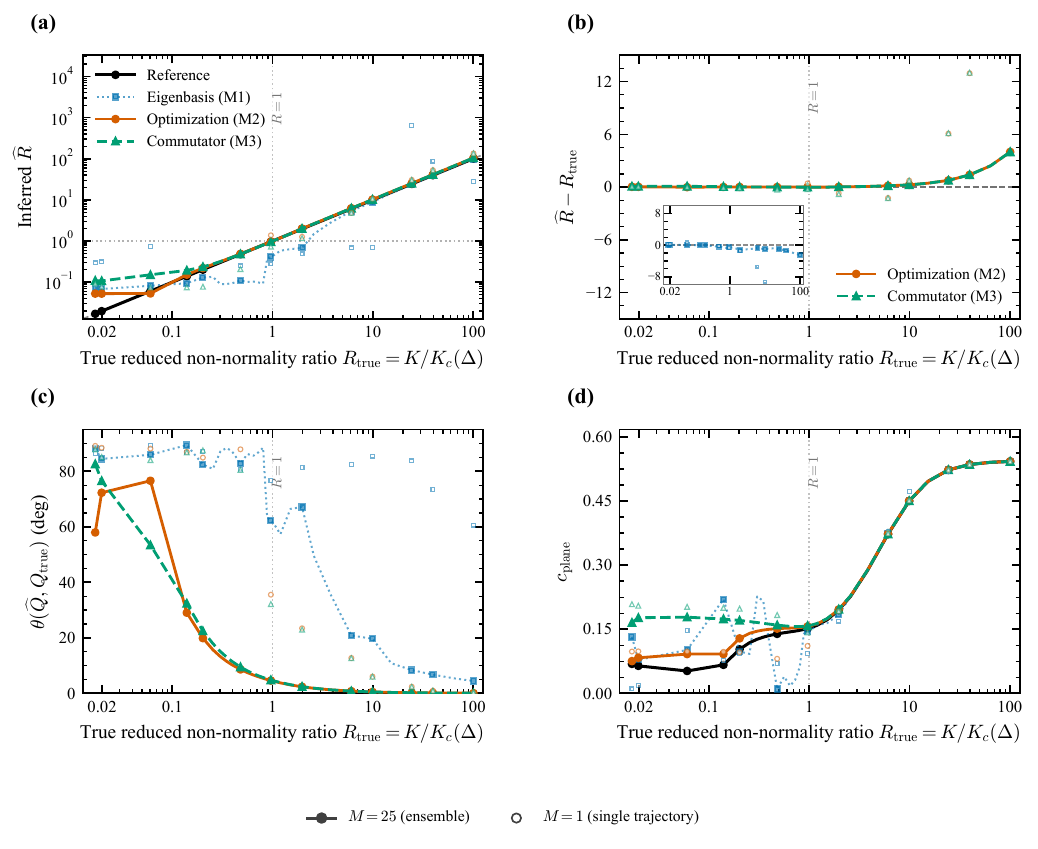}
    \caption{
    \textbf{Inferred reduced non-normality ratio \(R=K/K_c(\Delta)\) and response plane from estimated operators, for plane-extraction methods M1, M2, and M3.}
    Synthetic trajectories are generated from the VAR(1) model \(\x_{k+1}=F(\kappa)\x_k+\bm{\eta}_k\), with \(N=100\), \(\rho(F)=0.95\), \(\bm{\eta}_k\sim\mathcal{N}(0,\sigma^2I_N)\), and \(\sigma=0.05\). For each value of \(\kappa\), the fitted operator \(\Fhat\) is estimated by
concatenating \(M=25\) independent synthetic training trajectories, each of
length \(T_{\mathrm{train}}=300\), generated from the same operator \(F(\kappa)\). Filled markers connected by lines show this \(M=25\) ensemble estimate. Open markers show the corresponding \(M=1\) single-trajectory estimates, using the same method symbols. The \(M=1\) case matches the empirical moving-window setting, where each local operator is fitted from one observed trajectory segment.
The ridge parameter $\lambda$ is chosen by validation: for each candidate value in a logarithmic grid from $10^{-8}$ to $10^{2}$, $\widehat F$ is fitted on 80\% of the available training pairs and evaluated on the remaining 20\%; the value minimizing the validation one-step prediction error is retained. The test trajectory used for \(c_{\mathrm{plane}}\) (\ref{eq:cplane_def}) has length \(T_{\mathrm{test}}=1200\). The horizontal axis in all panels is the reference ratio \(R_{\mathrm{true}}=K_{\mathrm{true}}/K_c(\Delta_{\mathrm{true}})\), computed from the known operator \(F\). The vertical dotted line marks \(R_{\mathrm{true}}=1\).
    \textbf{(a)} Inferred ratio \(\widehat R\) for eigenbasis-SVD baseline extraction (M1), optimization-based directed-coupling extraction (M2), and commutator-based plane extraction (M3). The black diagonal curve is the reference value computed from \(F\).
    \textbf{(b)} Signed deviation \(\widehat R-R_{\mathrm{true}}\). The inset shows the same quantity for M1 on a larger vertical scale. Values outside the main vertical display range are indicated at the boundary; the full numerical deviations are retained in the data.
    \textbf{(c)} Plane recovery error \(\theta(Q,Q_{\mathrm{true}})\) (\ref{eq:principal_angle_def}), measured as the largest principal angle in degrees between the inferred plane and the reference plane.
    \textbf{(d)} Plane-invariant observable alignment \(c_{\mathrm{plane}}\) (\ref{eq:cplane_def}), computed between the empirical mean field and the best direction inside the inferred plane.
    }
    \label{fig:synthetic_method_comparison_trueR}
\end{figure*}

\Cref{fig:synthetic_method_comparison_trueR} compares the three plane-extraction methods after the operator has been estimated from finite data. In \cref{fig:synthetic_method_comparison_trueR}a, the estimated \(\widehat R\) obtained from optimization-based directed-coupling extraction (M2) and commutator-based plane extraction (M3) remain close to the true value over most of the parameter scan. The open \(M=1\) markers are noisier, as expected from a single finite trajectory, but they retain the same qualitative behavior of M2 and M3 near and above the reduced threshold. The eigenbasis-SVD baseline extraction (M1) underestimates \(\widehat R\) across a broad intermediate range. This shows that the main difficulty is not only estimating \(\Fhat\), but also extracting a stable response plane from the estimated operator.

The deviation $\widehat R - R_{\mathrm{true}}$ in \cref{fig:synthetic_method_comparison_trueR}b gives a more sensitive comparison of the scalar diagnostic. M2 and M3 exhibit small deviations relative to the range of \(R_{\mathrm{true}}\), whereas M1 has larger negative deviations, shown in the inset. These negative biases indicate that M1 should not be used as the primary estimator of the response plane, but only as a transparent baseline against which the directed M2 and M3 constructions can be compared.

The plane recovery error $\theta({\hat Q},Q_{\mathrm{true}})$ defined by expression (\ref{eq:principal_angle_def})
is shown in \cref{fig:synthetic_method_comparison_trueR}c. For \(R_{\mathrm{true}}\ll1\), the response plane is weakly expressed and is difficult to identify from finite trajectories. As \(R_{\mathrm{true}}\) approaches and exceeds the threshold \(R=1\), the principal-angle error for M2 and M3 decreases rapidly. M1 remains poorly aligned over much of the parameter range and improves only at the largest values of \(R_{\mathrm{true}}\). Thus the reduced response plane becomes identifiable from finite data when the non-normal geometry is sufficiently expressed.

Finally, \cref{fig:synthetic_method_comparison_trueR}d shows the plane-invariant correlation score $c_{\mathrm{plane}}$ defined by 
expression (\ref{eq:cplane_def}) as a function of $R$. In the strongly non-normal regime, the planes extracted by M2 and M3 contain directions whose activity is aligned with the empirical mean field. This is important for empirical applications, where the reference plane is not known and agreement between independent extraction methods becomes an internal consistency check.

% ------------------------------------------------------------
% ------------------------------------------------------------
\subsection{Dependence of reduced non-normal geometry recovery on number $M$ of samples, state dimension $N$, and training horizon $T$}
\label{sec:synthetic_scaling}

We next examine how recovery scales with the amount of data and with the ambient dimension.
To separate the relevant effects, we vary three quantities one at a time: 
the number $M$ of independent synthetic trajectories generated from the same operator, the state dimension $N$, 
and the training horizon $T_{\mathrm{train}}$. For each
setting, trajectories are generated from the controlled VAR(1) family, \(\Fhat\)
is estimated by ridge regression, and the three plane-extraction methods are
applied to the fitted operator. The scan over \(M\) includes \(M=1\), and the dimension and training-horizon scans include \(M=1\) insets for the intermediate regime \(R_{\mathrm{true}}\approx1\), to make the comparison with empirical single-segment fits explicit.

The reference plane is denoted by \(Q_{\mathrm{true}}\). Plane recovery is measured by the largest principal angle
$\theta(Q,Q_{\mathrm{true}})$ (\ref{eq:principal_angle_def}).  
Full-operator recovery is summarized by $\mathrm{relErr}(\Fhat)$ (\ref{eq:relerr_def}).
The accuracy of the estimated ratio is measured by
\begin{equation}
    \mathcal{E}_R
    =
    \left|
    \log_{10}
    \left(
    \frac{\widehat R}{R_{\mathrm{true}}}
    \right)
    \right| .
    \label{eq:ER_def}
\end{equation}
A value $\mathcal{E}_R=\log_{10}2$
corresponds to a factor-of-two error in the inferred ratio.

\begin{figure*}[!htpb]
    \centering
    \includegraphics[width=\textwidth]{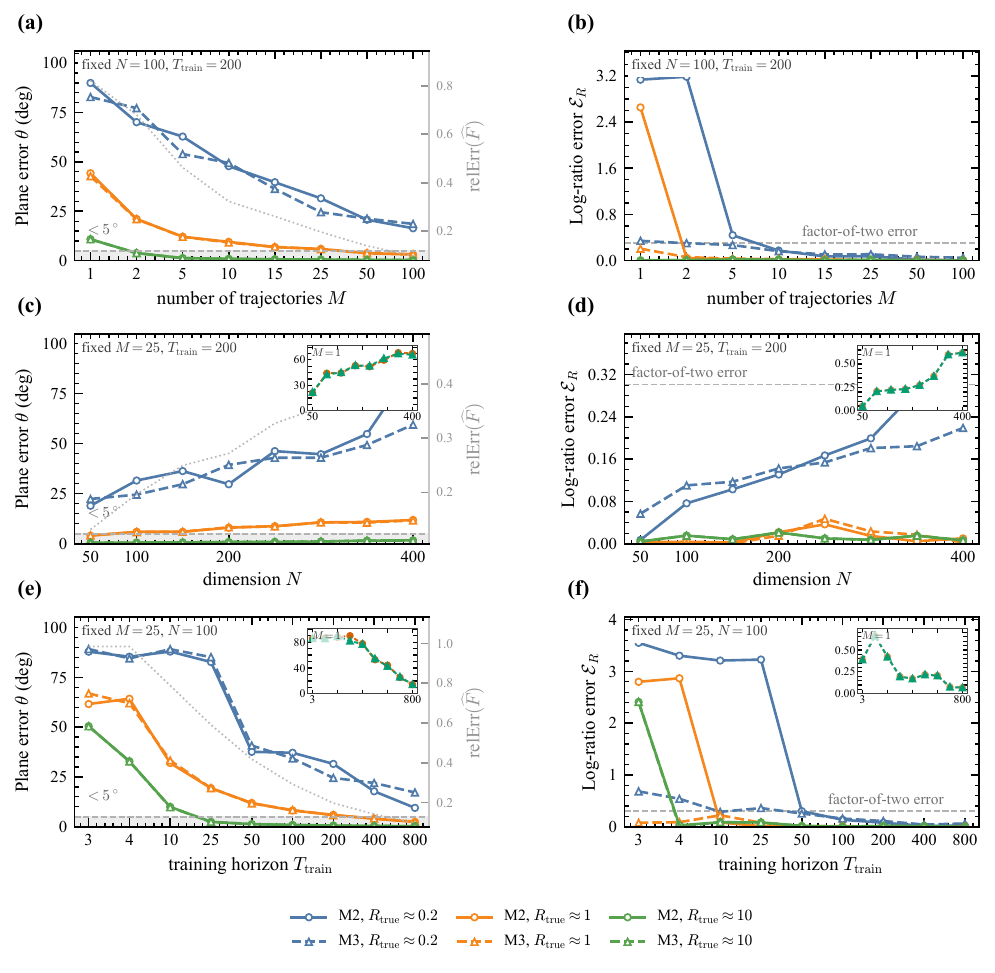}
    \caption{
    \textbf{Dependence of reduced non-normal geometry recovery on number $M$ of samples, state dimension $N$ and training horizon $T$.}
    Synthetic trajectories are generated from the VAR(1) model \(\x_{k+1}=F(\kappa)\x_k+\bm{\eta}_k\), with \(\rho(F)=0.95\), \(\bm{\eta}_k\sim\mathcal{N}(0,\sigma^2I_N)\), and \(\sigma=0.05\). For each scaling variable, the internal anisotropy parameter \(\kappa\) is chosen to obtain three reference regimes: \(R_{\mathrm{true}}\approx0.2\), \(R_{\mathrm{true}}\approx1\), and \(R_{\mathrm{true}}\approx10\), where \(R_{\mathrm{true}}=K_{\mathrm{true}}/K_c(\Delta_{\mathrm{true}})\). The left column shows the plane error \(\theta(Q,Q_{\mathrm{true}})\) in degrees. The right column shows the reduced-ratio error \(\mathcal{E}_R=\left|\log_{10}(\widehat R/R_{\mathrm{true}})\right|\). Solid curves denote optimization-based directed-coupling extraction (M2), and dashed curves denote commutator-based plane extraction (M3). The eigenbasis-SVD baseline extraction (M1) is  out-of-range and not shown. 
 In the left-column panels, the grey curve with scale corresponding to the right vertical axis shows the relative full-operator error $\mathrm{relErr}(\Fhat)$ (\ref{eq:relerr_def}) for the intermediate regime $R_{\mathrm{true}}\approx1$, providing a reference against which the reduced plane-recovery errors on the main axis can be compared.
    In the right-column panels, 
    the horizontal dashed line marks \(\mathcal{E}_R=\log_{10}2\), corresponding to a factor-of-two error.
    \textbf{(a,b)} Parameter scan over the number of trajectories \(M\in\{1,2,5,10,15,25,50,100\}\), with \(N=100\) and \(T_{\mathrm{train}}=200\).
    \textbf{(c,d)} Parameter scan  over the state dimension \(N\in\{50,100,150,200,250,300,350,400\}\), with \(M=25\) and \(T_{\mathrm{train}}=200\). Insets show the corresponding \(M=1\) single-trajectory results for \(R_{\mathrm{true}}\approx1\), using M2 and M3.
    \textbf{(e,f)} Parameter scan over the training horizon \(T_{\mathrm{train}}\in\{3,4,10,25,50,100,200,400,800\}\), with \(M=25\) and \(N=100\). Insets show the corresponding \(M=1\) single-trajectory results for \(R_{\mathrm{true}}\approx1\), using M2 and M3.
    }
    \label{fig:synthetic_scaling_trueR}
\end{figure*}

\Cref{fig:synthetic_scaling_trueR}a,b show the effect of increasing the number $M$ of independent trajectories. For fixed $N=100$ and $T_{\mathrm{train}}=200$, adding independent trajectories improves both operator estimation and reduced-geometry recovery. The leftmost point \(M=1\) gives the single-trajectory setting used later in the empirical moving-window analyses. The relative operator error decreases with $M$, while the plane errors of M2 and M3 decline and their reduced-ratio estimates eventually fall within the factor-of-two reference band. By contrast, the eigenbasis-SVD baseline extraction (M1) remains systematically less accurate; its errors fall outside the plotted range and are therefore omitted.

The dimension scan in \cref{fig:synthetic_scaling_trueR}c,d tests whether the reduced geometry remains identifiable as the ambient regression problem grows. With $M=25$ and $T_{\mathrm{train}}=200$ fixed, increasing $N$ makes the full $N\times N$ operator increasingly difficult to estimate entry by entry. Yet M2 and M3 preserve small plane errors and small reduced-ratio errors over the tested dimensions. The \(M=1\) insets show the same scan for the intermediate regime \(R_{\mathrm{true}}\approx1\); as expected, the single-trajectory errors are larger, but the M2 and M3 comparison remains visible. This shows that the dominant two-dimensional response geometry can remain identifiable even when full-matrix recovery becomes increasingly demanding.

The scan over training-horizons in \cref{fig:synthetic_scaling_trueR}e,f probes the regime of small data size. With \(M=25\) and \(N=100\) fixed, very short training horizons give larger errors because the regression has limited temporal information. As \(T_{\mathrm{train}}\) increases, both the plane error and the reduced-ratio error decrease for M2 and M3. The \(M=1\) insets provide the corresponding single-trajectory comparison for \(R_{\mathrm{true}}\approx1\). The M1 method leads to much larger errors, as expected from the sensitivity of eigenvector-based extraction to perturbations of the fitted operator.

Together, the three parameter scans show that the relevant object for the proposed reduction is not the full matrix \(\Fhat\) entry by entry, but the two-dimensional response plane and the ratio computed on that plane. The optimization-based and commutator-based methods recover these quantities across changes in \(M\), \(N\), and \(T_{\mathrm{train}}\), whereas the eigenbasis-SVD baseline is more fragile.

This data efficiency is worth emphasizing. Entrywise identification of a general \(N\times N\) linear operator requires on the order of \(N^2\) independent observations, since the operator has \(N^2\) free parameters. The scan over scaling shows that the reduced response plane and the ratio \(R\) are recovered well before this regime is reached: in the dimension scan, accurate recovery of both the response plane and $R$ is obtained with a fixed data budget of $M=25$ trajectories of length $T_{\mathrm{train}}=200$, even as $N$ increases to several hundred. The total number of observed time steps is therefore far below $N^2$, amounting to only a few percent of the number of entries in the full operator. The target of inference is not the full operator but a two-dimensional geometry, and this lower-dimensional target can be estimated from substantially fewer samples than full operator identification would demand.

% ------------------------------------------------------------
\subsection{Recovering the time-varying non-normal geometry ratio \(R=K/K_c(\Delta)\) with moving windows} 
\label{sec:synthetic_timevarying_revised}  

The preceding benchmarks used stationary operators. We now test whether the reduced diagnostic \(R\), defined in \eqref{eq:KKc_ratio_def}, can be tracked when the local dynamics changes over time. This experiment matches the moving-window setting used later for empirical recordings: for each selected window end time, a local operator is estimated from the data contained in that finite window.

\begin{figure*}[!htpb]     
\centering     
\includegraphics[width=\textwidth]{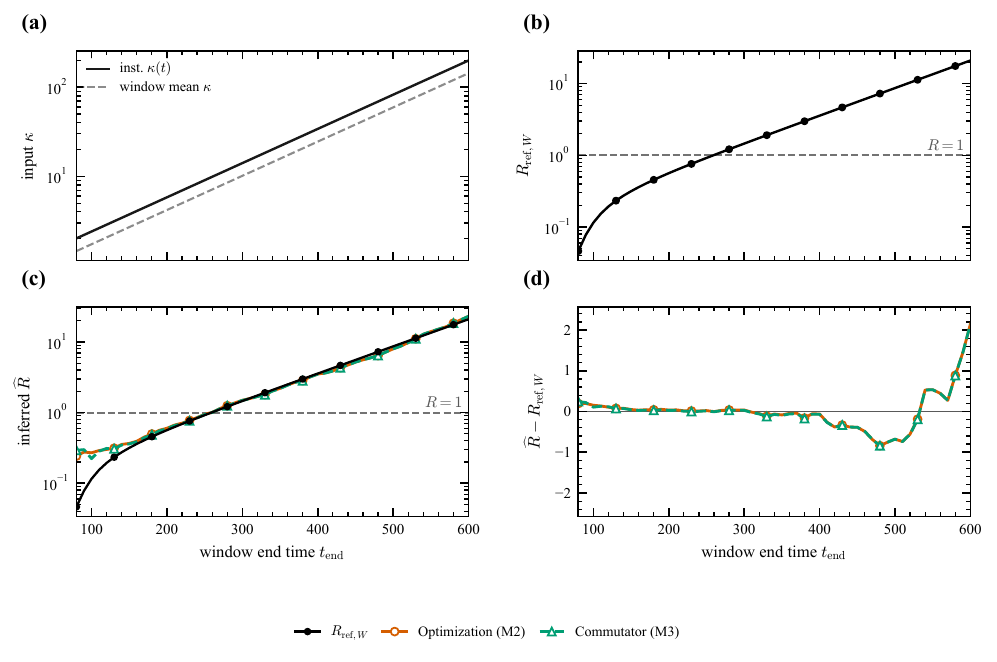}     
\caption{
Time-varying synthetic test with moving windows. Trajectories are generated from the time-varying VAR(1) model \eqref{eq:timevarying_var}, with \(F_t=F(\kappa_t)\), \(N=100\), \(M=1\), \(T=600\) time steps, \(\rho(F_t)=0.95\), \(\bm{\eta}_t\sim\mathcal{N}(0,\sigma^2I_N)\), and \(\sigma=0.05\). The parameter \(\kappa_t\) increases logarithmically from \(1\) to \(200\). Local operators are estimated using moving windows of length \(W=80\) with step size \(10\). Ridge regularization is selected from \(\lambda\in\{10^{-6},\ldots,10^1\}\) using an 80--20 train--validation split as in the previous synthetic tests.
\textbf{(a)} Instantaneous \(\kappa(t)\) and its moving-window mean.
\textbf{(b)} Window-matched reference ratio \(R_{\mathrm{ref},W}\), obtained by applying M2 to the window-averaged true operator \(\bar F_W(t_{\mathrm{end}})\) in \eqref{eq:Fbar_window}. The ratio \(R=K/K_c(\Delta)\) is defined in \eqref{eq:KKc_ratio_def}.
\textbf{(c)} Inferred ratio \(\widehat R\) from the fitted local operator \(\Fhat_W(t_{\mathrm{end}})\), compared with \(R_{\mathrm{ref},W}\). Orange circles correspond to M2 and green triangles to M3.
\textbf{(d)} Signed tracking error \(\widehat R-R_{\mathrm{ref},W}\) for M2 and M3. In panels~(b,c), the horizontal dashed line marks \(R=1\).
}
\label{fig:synthetic_timevarying_revised} 
\end{figure*}

We generate trajectories from a VAR(1) model \eqref{eq:synthetic_var1}
\begin{equation}     
\x_{t+1}
=
F_t\x_t+\bm{\eta}_t,
\qquad
F_t=F(\kappa_t),
\qquad
\bm{\eta}_t\sim\mathcal{N}(0,\sigma^2 I_N),
\label{eq:timevarying_var} 
\end{equation} 

where the internal anisotropy parameter \(\kappa_t\) increases logarithmically from \(1\) to \(200\) over the simulation interval \(t=1,\ldots,600\), as shown in \cref{fig:synthetic_timevarying_revised}. As before, \(\kappa_t\) is used only to construct the time-dependent operator \(F_t\). The reported diagnostic is the reduced ratio \(R=K/K_c(\Delta)\) defined in \eqref{eq:KKc_ratio_def}, computed from the reduced operator.

At each window end time \(t_{\mathrm{end}}\), advanced in steps of \(10\), a local operator is estimated from the moving window
\([t_{\mathrm{end}}-W,\ t_{\mathrm{end}})\), with window length \(W=80\). The fitted local operator is denoted by \(\Fhat_W(t_{\mathrm{end}})\). We use \(M=1\) in this test, matching the single-trajectory setting of the empirical moving-window analyses in \cref{sec:empirical}.

For each window, we compare \(\Fhat_W(t_{\mathrm{end}})\) with the window average of the true time-dependent operator over the same interval,
\begin{equation}    
\bar F_W(t_{\mathrm{end}})
=
\frac{1}{W}
\sum_{t=t_{\mathrm{end}}-W}^{t_{\mathrm{end}}-1}
F_t .
\label{eq:Fbar_window} 
\end{equation} 

Applying optimization-based directed-coupling extraction (M2) to \(\bar F_W(t_{\mathrm{end}})\) gives the window-matched reference value
\[
R_{\mathrm{ref},W}(t_{\mathrm{end}}).
\]
Applying M2 and commutator-based plane extraction (M3) to \(\Fhat_W(t_{\mathrm{end}})\) gives the inferred values \(\widehat R(t_{\mathrm{end}})\). The signed tracking error is
\[
\widehat R(t_{\mathrm{end}})-R_{\mathrm{ref},W}(t_{\mathrm{end}}).
\]
This comparison is window-matched: the inferred diagnostic is evaluated against the true operator averaged over the same interval used for estimation. With the right-endpoint labelling, no sample later than \(t_{\mathrm{end}}\) is used to estimate \(\Fhat_W(t_{\mathrm{end}})\).

\Cref{fig:synthetic_timevarying_revised} shows the moving-window estimate of the reduced ratio \(R\) as the local operator changes over time. Panel~(a) displays the imposed change in the internal construction parameter \(\kappa_t\) and its corresponding moving-window mean. Panel~(b) shows the window-matched reference \(R_{\mathrm{ref},W}\), which crosses \(R=1\) at approximately \(t_{\mathrm{end}}\approx260\) and then continues to increase. Panel~(c) compares this reference with the values inferred from fitted operators. The M2 and M3 curves remain close to the reference over most of the record. Panel~(d) shows that the signed error is small through the middle of the parameter scan and grows near the final windows, where the imposed change is largest on the scale of the moving window.

This test clarifies how the moving-window diagnostic should be interpreted in non-stationary data. The quantity reported at \(t_{\mathrm{end}}\) is not an instantaneous property of a single sample. It is a local diagnostic computed from the data in the moving window \([t_{\mathrm{end}}-W,t_{\mathrm{end}})\). The comparison with \(R_{\mathrm{ref},W}\) shows that this estimate follows the corresponding window-averaged reduced geometry. This supports the use of the same moving-window construction in the empirical analyses below.
% ------------------------------------------------------------
\subsection{Robustness to broad singular-value spectra}
\label{sec:synthetic_broad_singular_values}

The preceding synthetic benchmarks use controlled operator families in which the dominant response plane is known. We now ask whether the proposed two-dimensional reduction remains reliable when the surrounding high-dimensional operator has a broad singular-value spectrum. This test is important because, in empirical systems, the dominant non-normal response geometry need not appear as an isolated low-dimensional component against an otherwise featureless background. We therefore embed a prescribed two-dimensional non-normal block inside a stable high-dimensional background whose singular values span a broad hierarchy.

The full operator is defined as
\begin{equation} 
F = O \begin{pmatrix} B_R & 0\\ 0 & D_\beta \end{pmatrix} O^\top , 
\label{eq:broad_singular_operator} 
\end{equation}
where $O\in\mathbb{R}^{N\times N}$ is a random orthogonal matrix, $B_R\in\mathbb{R}^{2\times2}$ is a stable non-normal block with prescribed reference ratio $R_\star$, and $D_\beta$ is a stable diagonal background. The background is given by
\begin{equation}
D_\beta
=
\mathrm{diag}
\left(
c,1^{-\beta},
c,2^{-\beta},
\ldots,
c,(N-2)^{-\beta}
\right),
\label{eq}
\end{equation}
with $c=0.74$, so that the background remains stable and below the leading scale of the embedded block. The exponent $\beta$ controls the breadth of the background singular-value spectrum and is varied from $0.25$ to $2.25$. The embedded response plane is known by construction and is given by $Q_\star = O_{:,1:2}$.
We generate noisy trajectories from $F$, estimate $\widehat F$ by ridge regression, and apply optimization-based directed-coupling extraction M2 and commutator-based plane extraction M3 to the fitted operator. The goal is to determine whether the dominant two-dimensional non-normal geometry can still be recovered when it is embedded in a high-dimensional operator with substantial singular-value structure.

\begin{figure*}[!htpb]   
    \centering
    \includegraphics[width=\textwidth]{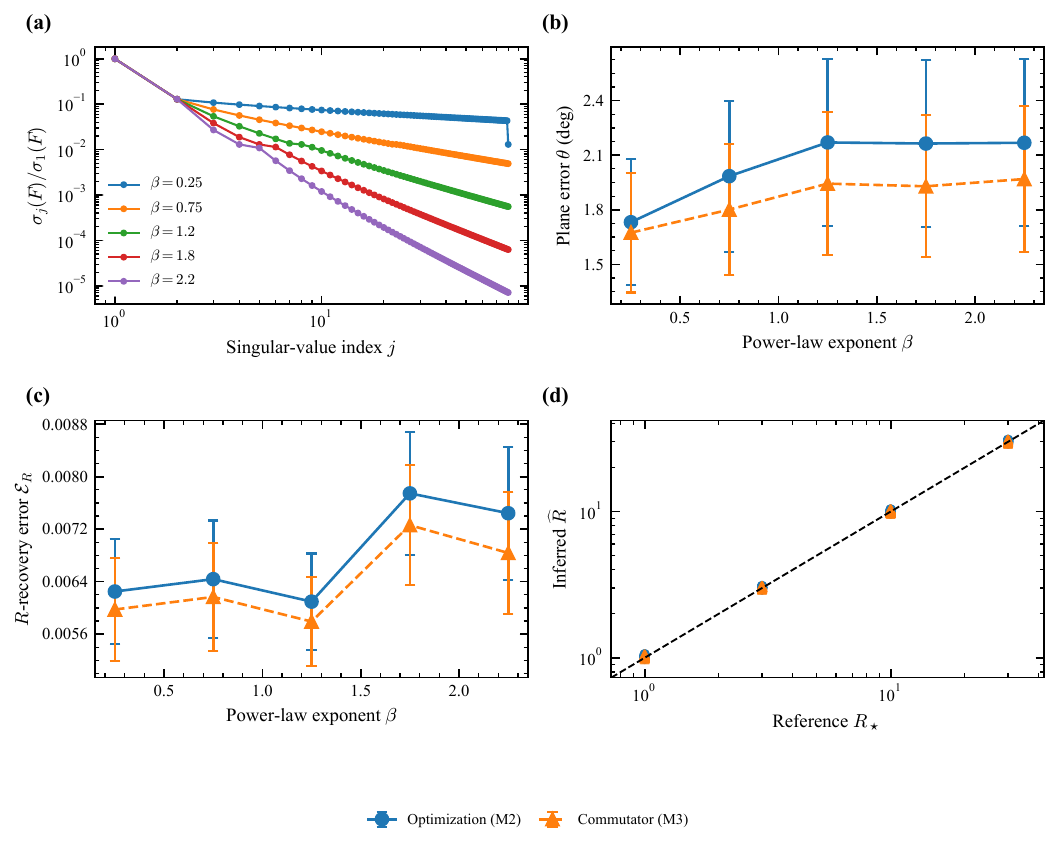}
    \caption{
    \textbf{Inferred response plane and ratio \(R=K/K_c(\Delta)\) for operators with broad singular-value spectra.}
    The full operator is \(F=O\,\mathrm{diag}(B_R,D_\beta)\,O^\top\) (\ref{eq:broad_singular_operator}), where \(O\) is random orthogonal, \(B_R\) is a stable two-dimensional non-normal block with prescribed reference ratio \(R_\star\), and \(D_\beta=\mathrm{diag}(c j^{-\beta})\) is a stable diagonal background with \(c=0.74\)
and exponent \(\beta\in\{0.25,0.75,1.25,1.75,2.25\}\). The benchmark uses \(N=80\), \(M=18\) independent training trajectories, \(T_{\mathrm{train}}=220\) samples per trajectory, additive Gaussian noise with \(\sigma=0.035\), ridge parameter \(\lambda=10^{-4}\), fixed embedded eigenvalue splitting \(\Delta=0.35\), reference ratios \(R_\star\in\{1,3,10,30\}\) and 8 random realizations.
    \textbf{(a)} Normalized singular-value spectra \(\sigma_j(F)/\sigma_1(F)\) for the five values of \(\beta\), shown for \(R_\star=3\).
    \textbf{(b)} Principal-angle error \(\theta(\widehat Q,Q_\star)\) (\ref{eq:principal_angle_def},\ref{eq:principal_angle_def}), where $Q_\star = O_{:,1:2}$, between the inferred plane and the embedded response plane, plotted as mean \(\pm\) standard error over all \(R_\star\) values and random realizations for each \(\beta\).
    \textbf{(c)} Error \(\mathcal E_R=\left|\log_{10}(\widehat R/R_\star)\right|\), plotted as mean \(\pm\) standard error over all \(R_\star\) values and random realizations for each \(\beta\).
    \textbf{(d)} Inferred ratio \(\widehat R\) versus reference ratio \(R_\star\) for all \(\beta\), \(R_\star\), random realizations, and both extraction methods. The dashed line indicates \(\widehat R=R_\star\). In panels~(b--d), circles refer to optimization-based directed-coupling extraction (M2), and triangles refer to commutator-based plane extraction (M3).
    }
    \label{fig:broad_singular_value_nndr}
\end{figure*}

\Cref{fig:broad_singular_value_nndr} tests whether the proposed reduction can identify a prescribed non-normal response geometry when it is embedded in a high-dimensional operator with a nontrivial singular-value hierarchy. Panel~(a) shows the range of background singular-value structures used in the test. For large $\beta$, the singular values span more than five orders of magnitude, so the leading two-dimensional structure is relatively well separated. For small $\beta$, in particular $\beta=0.25$, the spectrum spans less than two decades, leaving many background directions of comparable scale and making recovery of the embedded response plane a more stringent test. The remaining panels then ask whether this additional singular structure interferes with recovery of the embedded response plane $Q_\star=O_{:,1:2}$ and of its prescribed reduced ratio $R_\star$.

Panels~(b) and~(c) show that both optimization-based directed-coupling extraction (M2) and commutator-based plane extraction (M3) obtain small principal-angle errors and small reduced-ratio errors across the tested values of $\beta$. Thus, recovery is not restricted to an artificially simple operator whose singular spectrum contains only one dominant low-dimensional component. Panel~(d) pools all values of $\beta$, all reference ratios $R_\star\in{1,3,10,30}$, all random realizations, and both extraction methods. The inferred ratios remain close to the identity line $\widehat R=R_\star$, showing that the scalar diagnostic is preserved not only near threshold but also in strongly non-normal regimes.

This benchmark clarifies what the two-dimensional projection is meant to do. It is not a rank-two approximation to the full operator and does not require the remaining $N-2$ directions to be dynamically negligible. Rather, it seeks the two-dimensional subspace in which the input--response geometry responsible for transient amplification is concentrated. The results show that this geometry remains identifiable even when the full operator contains many additional singular directions.

% ============================================================
\section{Empirical applications of non-normal response inference}
\label{sec:empirical}

\subsection{From synthetic validation to empirical moving-window diagnostics}
\label{sec:empirical_overview}

The synthetic benchmarks in Section~5 show that the reduced response plane and the reduced ratio \(R\) defined in \eqref{eq:KKc_ratio_def} can be recovered from time series. There, a reference operator and response plane were available, but only for scoring; the inference itself never used them.

We now apply the same construction to empirical recordings, where the local operator and reference plane are genuinely unavailable. These examples are therefore demonstrations, not ground-truth tests. We read them through one standard, stated here and used throughout the section, so that the individual subsections need only report their results. An increase in \(R\) means the fitted local reduction has moved toward or past the two-dimensional threshold \(R=1\) of \eqref{eq:KKc_ratio_def}. We treat such a change as event-associated only when four conditions hold together: M2 and M3 agree in time, the diagnostics are coherent across neighboring windows rather than isolated spikes, the change aligns with an independently defined event or activity reference, and \(\Delta\) stays away from the degenerate limit. These are the checks each subsection applies. None of the examples is a detector, and \(R=1\) is not a clinical or behavioral decision boundary.

We consider four settings: electrohysterogram (EHG) recordings of uterine electrical activity, epileptic seizure EEG, freezing-of-gait acceleration, and wearable inertial data from rhythmic push-ups. They differ in sensor modality, sampling rate, preprocessing, and event definition, so window lengths, filtering, regularization, and alignment rules are set per dataset. The procedure is otherwise common: estimate a local operator from a moving window, extract a two-dimensional response plane, and compute \(R\) \eqref{eq:KKc_ratio_def}, the support \(S(\rhat)\) \eqref{eq:empirical_support} and \(\Delta\) \eqref{eq:Delta_def}. We present the settings in order of decreasing evidential strength, from the multi-level uterine analysis to the single-record push-up example, and summarize them in \cref{tab:empirical_overview}.

% ------------------------------------------------------------
\begin{table*}[!htbp]
\centering
\footnotesize
\caption{\textbf{Overview of the four empirical applications of non-normal response inference.} For each dataset, we list the multivariate state, the moving-window length and step, the independently defined event or activity reference, the change observed in the reduced diagnostics \(R\) and \(\Delta\), and the level of evidence. The quantities \(R\) and \(\Delta\) are defined in \eqref{eq:KKc_ratio_def} and \eqref{eq:Delta_def}. The settings are ordered by decreasing strength of evidence.}
\label{tab:empirical_overview}
\begin{tabular}{@{}p{2.5cm}p{2.2cm}p{1.9cm}p{3.0cm}p{3.4cm}p{2.0cm}@{}}
\toprule
Dataset & State (channels) & Window / step & Event or activity reference & Observed change & Evidence level \\
\midrule
Uterine EHG (parturition) & multichannel EHG array & \(60\,\mathrm{s}\) / \(5\,\mathrm{s}\) & elevated EHG activity envelope & \(R\) rises with activity; broadly supported reaction direction & strongest: representative, peak-aligned, and 45-subject analyses \\
Seizure EEG & scalp EEG (CHB-MIT) & \(40\,\mathrm{s}\) / \(2\,\mathrm{s}\) & annotated seizure onset & onset-aligned change in \(R\) and \(\Delta\) & representative and patient-pooled cohort \\
Freezing of gait & 3 ankle acceleration axes &  \(3\,\mathrm{s}\) / \(0.5\,\mathrm{s}\) & annotated freezing onset & \(R\) rises at the onset interval & representative and dataset-level onset vs baseline \\
Push-up motion & 3 acceleration axes & high- vs low-acceleration & acceleration-amplitude envelope &high-acceleration \(R\) larger than low but below \(R=1\); reaction coordinate aligned with high-acceleration intervals & directional only; subthreshold \\
\bottomrule
\end{tabular}
\end{table*}

\subsection{Moving-window inference in empirical data}
\label{sec:empirical_rolling_window}

For each dataset, we construct a multivariate state
\begin{equation}
    \x(t)
    =
    \big[x_1(t),x_2(t),\ldots,x_N(t)\big]^\top ,
    \label{eq:empirical_state}
\end{equation}
whose components are channels, sensors, or embedded coordinates, depending on the application. Each recording is first preprocessed for its modality: the filtering, downsampling, demeaning, standardization, and event alignment given in the corresponding subsection.

The data are then analyzed in moving windows. The \(j\)-th window holds the samples
\begin{equation}
    \mathcal{W}_j
    =
    \{t_j,t_j+1,\ldots,t_j+W-1\},
    \label{eq:empirical_window}
\end{equation}
where \(W\) is the window length in samples after preprocessing, and consecutive windows are separated by a dataset-specific step. Each window is labelled by its right endpoint,
\begin{equation}
    t_{\mathrm{end},j}
    =
    t_j+W .
    \label{eq:empirical_window_end}
\end{equation}
The diagnostic displayed at \(t_{\mathrm{end},j}\) is computed from the samples in \(\mathcal{W}_j\), that is from the interval \([t_{\mathrm{end},j}-W,\,t_{\mathrm{end},j})\); it is not a property of the single sample \(t_{\mathrm{end},j}\). This is the endpoint convention of the time-varying synthetic test in \cref{sec:synthetic_timevarying_revised}. Each window supplies one consecutive trajectory segment, so the empirical setting corresponds to \(M=1\) in the notations used for the synthetic tests.

In each window, we fit a local one-step model
\begin{equation}
    \x_{k+1}
    \approx
    \Fhat_j \x_k ,
    \qquad
    k=t_j,\ldots,t_j+W-2 ,
    \label{eq:empirical_local_fit}
\end{equation}
by ridge-regularized regression. When a continuous-time generator is estimated by finite differences, the same reduction is applied to that generator; we write \(\Fhat_j\) for the fitted local operator in either case.

Short windows can give poorly conditioned regressions or nearly degenerate reduced matrices, and either can inflate \(R\) for numerical rather than physical reasons. We therefore apply mild stabilization and screening. The stabilized operator is
\begin{equation}
    \Fhat_{j,\mathrm{reg}}
    =
    (1-\alpha)\Fhat_j+\alpha \mu_j I_N,
    \qquad
    \mu_j=\frac{\tr(\Fhat_j)}{N},
    \label{eq:empirical_shrinkage}
\end{equation}
with \(\alpha\) a dataset-specific shrinkage level. A window is excluded when the regression is too poorly conditioned, when the fitted operator is numerically unstable beyond what regularization repairs, or when the reduced \(2\times2\) matrix is too degenerate for a meaningful \(\Delta\) in \eqref{eq:Delta_def} and \(K_c(\Delta)\) in \eqref{eq:Kc_def}. Excluded windows are missing values, not extreme values of \(R\).

For each accepted window, optimization-based directed-coupling extraction (M2), described in \cref{sec:method2}, and commutator-based plane extraction (M3), described in \cref{sec:method3}, give
\begin{equation}
    Q_j^m
    =
    [\rhat_j^m,\nhat_j^m],
    \qquad
    m\in\{\mathrm{M2},\mathrm{M3}\}.
    \label{eq:empirical_Q_methods}
\end{equation}
For M3, the internal ordering step of \cref{sec:method3} assigns the response and input directions. The reduced operator is
\begin{equation}
    \Gam_j^m
    =
    (Q_j^m)^\top
    \Fhat_{j,\mathrm{reg}}
    Q_j^m,
    \qquad
    m\in\{\mathrm{M2},\mathrm{M3}\},
    \label{eq:empirical_Gamma_methods}
\end{equation}
and from it we compute
\begin{equation}
    \mathcal{D}_j^m
    =
    \{\Delta_j^m,\kappa_{2D,j}^m,K_j^m,R_j^m\},
    \qquad
    m\in\{\mathrm{M2},\mathrm{M3}\},
    \label{eq:empirical_reduced_quantities}
\end{equation}
using the definitions \eqref{eq:Delta_def}--\eqref{eq:KKc_ratio_def} with \(\Gam\) replaced by \(\Gam_j^m\). The level \(R_j^m=1\) is the reference threshold from the reduced calculation, not a clinical or behavioral boundary.

When an observable signal is available, we compare it with the reaction coordinate
\begin{equation}
    y_{r,j}^m(t)
    =
    (\rhat_j^m)^\top \x(t),
    \qquad
    t\in\mathcal{W}_j,
    \qquad
    m\in\{\mathrm{M2},\mathrm{M3}\}.
    \label{eq:empirical_reaction_projection}
\end{equation}
The observable is dataset-specific: an EHG activity envelope, EEG mean-field activity, acceleration magnitude, or an acceleration-amplitude envelope. We also report the reaction support, which separates a response direction spread across many channels from one concentrated on a few. For a normalized reaction direction \(\rhat_j^m\), the reaction support reads
\begin{equation}
    S(\rhat_j^m)
    =
    \frac{\|\rhat_j^m\|_1}{\sqrt{N}\|\rhat_j^m\|_2},
    \qquad
    0<S(\rhat_j^m)\leq 1 .
    \label{eq:empirical_support}
\end{equation}
Values near one mean the response direction is distributed across many components; smaller values mean it is localized.

The following subsections apply this construction to the four datasets. In the EHG, seizure, and freezing-of-gait analyses, we ask whether an independently defined event coincides with changes in \(R\), \(S(\rhat)\), and \(\Delta\). The push-up example asks a different question: whether the high-acceleration samples carry a larger \(R\) than the low-acceleration samples, even while \(R\) stays below \(1\).

% ------------------------------------------------------------
\subsection{Parturition and uterine electrical activity}
\label{sec:ehg}

Uterine contractions are coordinated physiological events in which electrical activation spreads across the abdominal electrode array. Multichannel electrohysterogram (EHG) is therefore a natural first test of the method: the activation is spatially distributed, so a coherent reduced response plane, if one exists, should be visible across channels. We ask whether periods of elevated EHG-derived activity coincide with changes in the reduced ratio \(R\) defined in \eqref{eq:KKc_ratio_def}, the reaction support \(S(\rhat)\) defined in \eqref{eq:empirical_support}, and the reduced eigenvalue splitting \(\Delta\) defined in \eqref{eq:Delta_def}.

The analysis used all available EHG channels. The raw signals were cropped to remove the initial contaminated segment and end effects, converted from \(\mathrm{mV}\) to \(\mu\mathrm{V}\), bandpass filtered in the uterine band \(0.08\text{--}2.0\,\mathrm{Hz}\), and downsampled to \(10\,\mathrm{Hz}\). Local generators were estimated on \(60\,\mathrm{s}\) moving windows advanced in \(5\,\mathrm{s}\) steps by ridge-regularized least squares. Each diagnostic is plotted at the window end time \eqref{eq:empirical_window_end}. The fitted generator \(\Fhat_j\) was stabilized by the trace-preserving shrinkage in \eqref{eq:empirical_shrinkage} at level \(\alpha=0.1\), because freely moving subjects can produce short nonstationary bursts that destabilize the local fit. The reduced plane was extracted independently by optimization-based directed-coupling extraction (M2) and commutator-based plane extraction (M3). For each accepted window, we computed the reduced quantities in \eqref{eq:empirical_reduced_quantities} and the support in \eqref{eq:empirical_support}. Windows rejected by the screening rules in \cref{sec:empirical_rolling_window} were treated as missing values.

For an activity reference, we used the EHG signal itself. Let \(x_i(t)\) denote the filtered EHG on channel \(i\), and let
\begin{equation}
    z_i(t)
    =
    x_i(t)
    +
    \mathrm{i}\,\mathcal{H}[x_i](t)
    \label{eq:ehg_analytic_signal}
\end{equation}
be its analytic signal, where \(\mathcal{H}\) is the Hilbert transform. The EHG-derived activity envelope is
\begin{equation}
    s_{\mathrm{EHG}}(t)
    =
    \frac{1}{N}
    \sum_{i=1}^{N}
    |z_i(t)|.
    \label{eq:ehg_activity_envelope}
\end{equation}
This is an electrical envelope; it is not a measurement of mechanical contraction force.

We also asked whether the reaction coordinate accounts for the coherent EHG mean field. The reaction coordinate \(y_{r,j}^m(t)\) is defined in \eqref{eq:empirical_reaction_projection}, and the mean field is
\begin{equation}
    m(t)
    =
    \frac{1}{N}
    \sum_{i=1}^{N}x_i(t).
    \label{eq:ehg_meanfield}
\end{equation}
For each accepted window and each method \(m\in\{\mathrm{M2},\mathrm{M3}\}\), we measured
\begin{equation}
    R^2_{\rhat\to m,j}
    =
    \mathrm{corr}^{2}
    \left(
        y_{r,j}^m(t),
        m(t)
    \right),
    \qquad
    t\in\mathcal{W}_j ,
    \label{eq:ehg_reaction_meanfield_r2}
\end{equation}
the fraction of the mean-field fluctuations reconstructed by the scalar reaction coordinate. The superscript $^2$ in \eqref{eq:ehg_reaction_meanfield_r2} indicates that 
$R^2_{\rhat\to m,j}$ a coefficient of determination; it is not the reduced ratio \(R\) of \eqref{eq:KKc_ratio_def}. We also computed the alignment between the reaction direction and the uniform mode,
\begin{equation}
    A_{\rhat,j}^{m}
    =
    \left|
    \left\langle
    \rhat_j^m,
    \frac{\bm{1}}{\sqrt{N}}
    \right\rangle
    \right|,
    \qquad
    m\in\{\mathrm{M2},\mathrm{M3}\}.
    \label{eq:ehg_meanfield_alignment}
\end{equation}
A large \(R^2_{\rhat\to m,j}\) means the reaction coordinate reconstructs the coherent mean-field fluctuation; a large \(A_{\rhat,j}^{m}\) means the reaction direction \(\rhat_j^m\) is itself close to the uniform mode.

\begin{figure*}[!htpb]
    \centering
    \includegraphics[height=0.74\textheight,keepaspectratio]{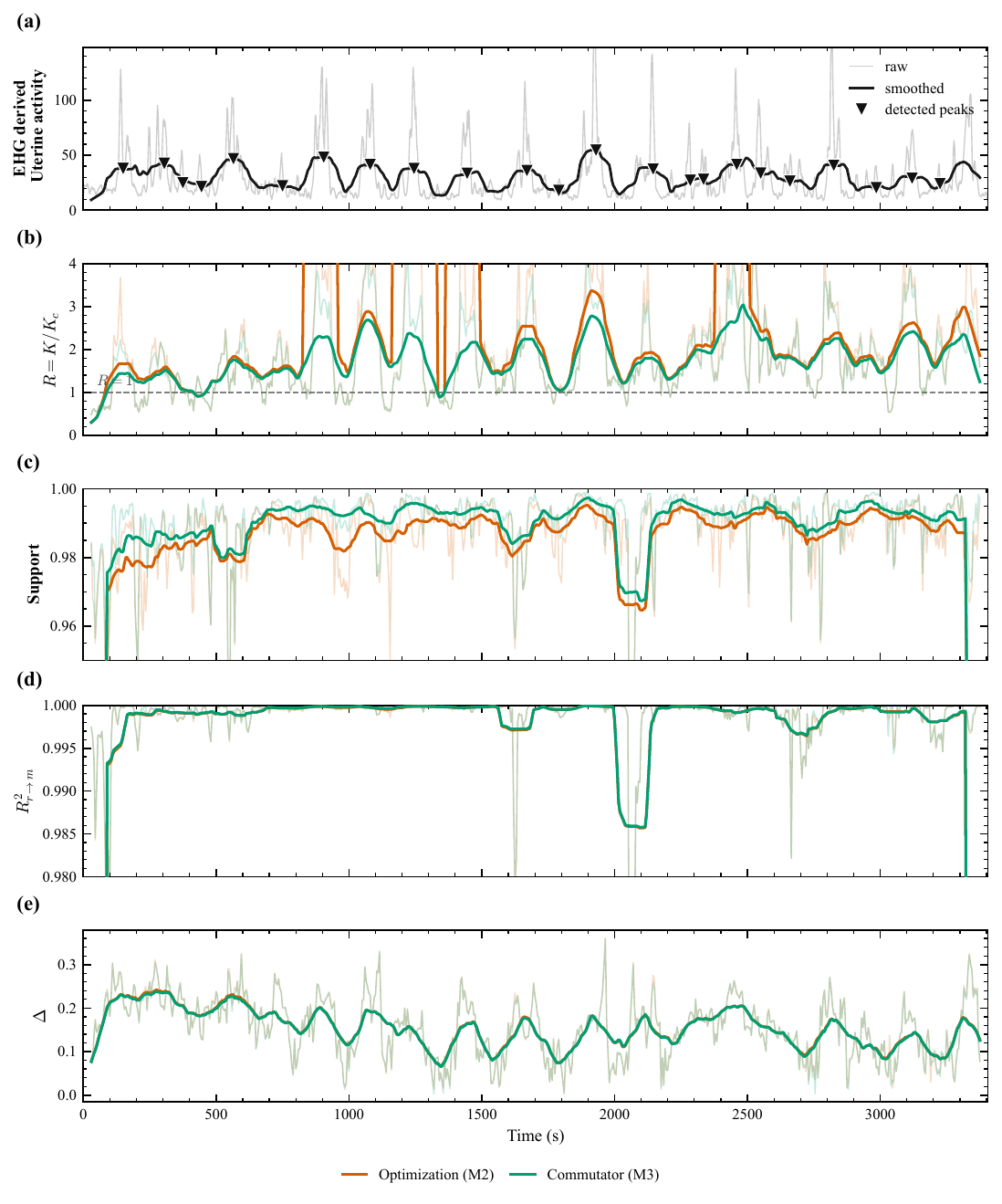}
    \caption{
    \textbf{Moving-window EHG diagnostics in one representative recording.}
    One multichannel EHG recording analyzed with \(60\,\mathrm{s}\) moving windows advanced in \(5\,\mathrm{s}\) steps. Diagnostics are plotted at the window end time \eqref{eq:empirical_window_end}. The signals were bandpass filtered between \(0.08\) and \(2.0\,\mathrm{Hz}\), converted to \(\mu\mathrm{V}\), and downsampled to \(10\,\mathrm{Hz}\). Local generators were estimated by ridge-regularized least squares and stabilized by the shrinkage in \eqref{eq:empirical_shrinkage} with
    \(\alpha=0.1\). Orange curves denote optimization-based directed-coupling extraction (M2), and green curves denote commutator-based plane extraction (M3). Thin transparent curves show unsmoothed values, thick curves show smoothed versions.
    \textbf{(a)} EHG-derived activity envelope \(s_{\mathrm{EHG}}(t)\) defined in \eqref{eq:ehg_activity_envelope}. The grey curve is the raw envelope, the black curve a \(120\,\mathrm{s}\) moving average, and inverted triangles mark detected activity peaks.
    \textbf{(b)} Reduced ratio \(R\) defined in \eqref{eq:KKc_ratio_def}. The dashed horizontal line marks \(R=1\).
    \textbf{(c)} Reaction support \(S(\rhat)\) defined in \eqref{eq:empirical_support}.
    \textbf{(d)} Reaction--mean-field coefficient of determination \(R^2_{\rhat\to m,j}\) defined in \eqref{eq:ehg_reaction_meanfield_r2}.
    \textbf{(e)} Reduced eigenvalue splitting \(\Delta\) defined in \eqref{eq:Delta_def}.
    }
    \label{fig:ehg_representative}
\end{figure*}

\Cref{fig:ehg_representative} shows the diagnostics for one representative recording. The activity envelope \eqref{eq:ehg_activity_envelope} in \cref{fig:ehg_representative}a is the reference. The reduced ratio \(R\) in \cref{fig:ehg_representative}b rises repeatedly during broad episodes of elevated activity, and the M2 and M3 curves share the same temporal structure. The support \(S(\rhat)\) in \cref{fig:ehg_representative}c stays near one over most windows, so the reaction direction is spread across the electrode array rather than concentrated on a few channels. The reaction--mean-field \(R^2_{\rhat\to m,j}\) in \cref{fig:ehg_representative}d is also near one for most of the recording: the reaction coordinate captures the coherent mean-field component. The splitting \(\Delta\) in \cref{fig:ehg_representative}e stays finite and smooth, so the rises in \(R\) are not numerical singularities of the reduced matrix.

\begin{figure}[!htpb]
    \centering
    \includegraphics[width=\columnwidth]{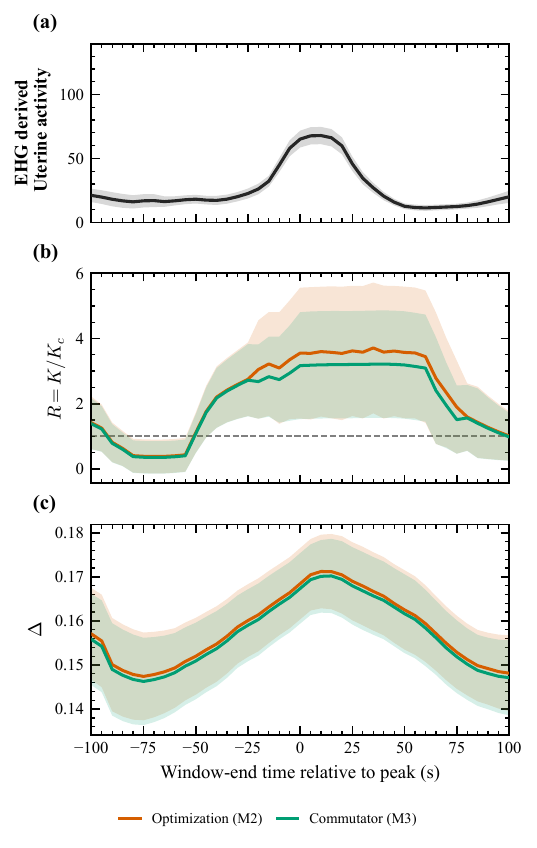}
    \caption{
    \textbf{EHG diagnostics aligned to detected activity peaks in the representative recording.}
    The activity peaks detected in \cref{fig:ehg_representative} are aligned at time zero, using 23 peaks and a \([-100,100]\,\mathrm{s}\) window around each. Orange curves denote optimization-based directed-coupling extraction (M2), and green curves denote commutator-based plane extraction (M3). Shaded bands show the across-peak variability.
    \textbf{(a)} Peak-aligned activity envelope \(s_{\mathrm{EHG}}(t)\) defined in \eqref{eq:ehg_activity_envelope}.
    \textbf{(b)} Peak-aligned reduced ratio \(R\) defined in \eqref{eq:KKc_ratio_def}. The dashed horizontal line marks \(R=1\).
    \textbf{(c)} Peak-aligned reduced eigenvalue splitting \(\Delta\) defined in \eqref{eq:Delta_def}.
    The vertical dotted line marks the activity peak.
    }
    \label{fig:ehg_peak_aligned}
\end{figure}

The peak-aligned analysis in \cref{fig:ehg_peak_aligned} tests whether one isolated episode drives the recording or whether the pattern repeats. With the detected peaks aligned, the envelope has a clear maximum at the alignment time, \(R\) is elevated over the same interval for both M2 and M3, and \(\Delta\) varies smoothly across the peak. The pattern therefore recurs across the detected peaks within the recording rather than reflecting a single episode.

To test whether the association holds across the dataset, we repeated the analysis on \(123\) recordings from \(45\) subjects. Because the envelope amplitude is recording-dependent, we compared high-activity and baseline windows within each recording. For recording \(j\), let \(s_j(t_{\mathrm{end}})\) be the smoothed envelope on the endpoint grid, and define
\begin{align}
    \mathcal{H}_j
    &=
    \{t_{\mathrm{end}}: s_j(t_{\mathrm{end}})\ge Q_{0.80}[s_j]\}, \nonumber \\
    \mathcal{B}_j
    &=
    \{t_{\mathrm{end}}: s_j(t_{\mathrm{end}})\le Q_{0.50}[s_j]\},
    \label{eq:ehg_high_baseline_sets}
\end{align}
where \(Q_p[s_j]\) is the \(p\)-quantile within recording \(j\). Thus \(\mathcal{H}_j\) holds the top \(20\%\) high-activity windows and \(\mathcal{B}_j\) the bottom \(50\%\). For any diagnostic \(Y_j^m(t_{\mathrm{end}})\), the high-minus-baseline difference is
\begin{equation}
\begin{aligned}
    d_j^m(Y)
    &=
    \mathrm{median}_{t_{\mathrm{end}}\in\mathcal{H}_j}
    Y_j^m(t_{\mathrm{end}}) \\
    &\quad -
    \mathrm{median}_{t_{\mathrm{end}}\in\mathcal{B}_j}
    Y_j^m(t_{\mathrm{end}}),
    \qquad
    m\in\{\mathrm{M2},\mathrm{M3}\}.
\end{aligned}
\label{eq:ehg_high_baseline_difference}
\end{equation}
For \(Y=R\), with \(R\) defined in \eqref{eq:KKc_ratio_def}, we also compute the per-recording Spearman correlation
\begin{equation}
    \rho_j^m
    =
    \rho_{\mathrm{S}}
    \left(
        s_j(t_{\mathrm{end}}),
        R_j^m(t_{\mathrm{end}})
    \right),
    \qquad
    m\in\{\mathrm{M2},\mathrm{M3}\}~.
    \label{eq:ehg_spearman}
\end{equation}
Cliff's delta, $\delta=\mathbb{P}(R_{\mathcal{H}j}>R{\mathcal{B}j})-\mathbb{P}(R{\mathcal{H}j}<R{\mathcal{B}j})$, was used as a non-parametric effect size to compare $R_j^m(t{\mathrm{end}})$ between $\mathcal{H}_j$ and $\mathcal{B}_j$. For each subject, recording-level quantities were summarized by their median, so that each subject contributed a single value.

\begin{figure*}[!htpb]
    \centering
    \includegraphics[width=\textwidth]{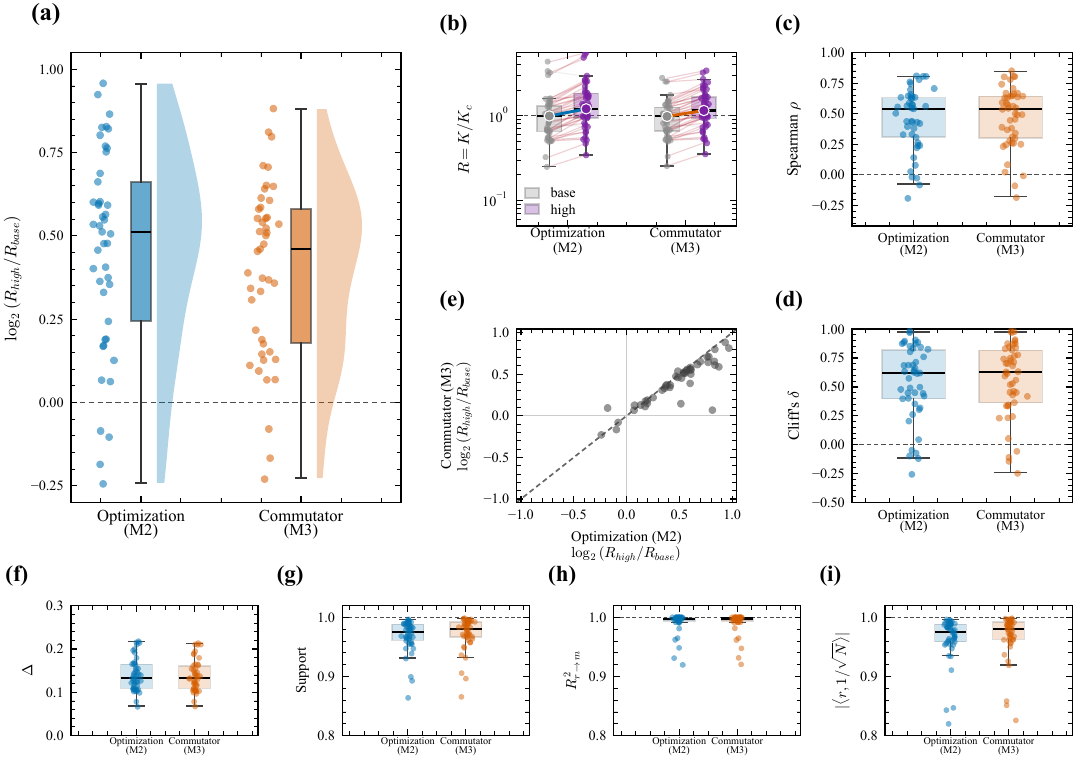}
    \caption{
    \textbf{Subject-level summary of EHG diagnostics during high-activity and baseline windows.}
    The analysis includes \(123\) recordings from \(45\) subjects. High-activity windows are the top \(20\%\) of the smoothed activity envelope within each recording, baseline windows the bottom \(50\%\), as defined in \eqref{eq:ehg_high_baseline_sets}. Recording-level quantities are summarized within each subject by their median. Blue denotes optimization-based directed-coupling extraction (M2), orange denotes commutator-based plane extraction (M3).
    \textbf{(a)} Distribution of \(\log_2(R_{\mathrm{high}}/R_{\mathrm{base}})\), where \(R_{\mathrm{high}}\) and \(R_{\mathrm{base}}\) are median values of \(R\) \eqref{eq:KKc_ratio_def} in high-activity and baseline windows.
    \textbf{(b)} Paired subject-level values of \(R_{\mathrm{base}}\) and \(R_{\mathrm{high}}\).
    \textbf{(c)} Spearman correlation \(\rho_j^m\) between the activity envelope and \(R\), defined in \eqref{eq:ehg_spearman}.
    \textbf{(d)} Cliff's delta comparing \(R\) between high-activity and baseline windows.
    \textbf{(e)} Method agreement between M2 and M3 for \(\log_2(R_{\mathrm{high}}/R_{\mathrm{base}})\). The dashed diagonal marks equality.
    \textbf{(f)} Subject-level \(\Delta\) defined in \eqref{eq:Delta_def} in high-activity windows.
    \textbf{(g)} Reaction support \(S(\rhat)\) defined in \eqref{eq:empirical_support} in high-activity windows.
    \textbf{(h)} Reaction--mean-field coefficient of determination \(R^2_{\rhat\to m,j}\) defined in \eqref{eq:ehg_reaction_meanfield_r2} in high-activity windows.
    \textbf{(i)} Alignment \(A_{\rhat,j}^{m}\) between the reaction direction and the uniform mode, defined in \eqref{eq:ehg_meanfield_alignment}, in high-activity windows.
    }
    \label{fig:ehg_dataset}
\end{figure*}

\Cref{fig:ehg_dataset} summarizes the association across subjects. Panels~(a) and~(b) show larger \(R\) in high-activity than in baseline windows for both M2 and M3. Panel~(c) shows a positive within-recording rank correlation between the activity envelope and \(R\), and panel~(d) the corresponding Cliff's delta. Panel~(e) shows that the \(\log_2(R_{\mathrm{high}}/R_{\mathrm{base}})\) values from M2 and M3 agree across subjects.
Panels (f)--(i) address whether the association is geometrically interpretable. \(\Delta\) stays away from the degenerate limit for most subjects, so the reduced matrix is well defined. The support \(S(\rhat)\) stays high, so the reaction direction remains spread across the array. \(R^2_{\rhat\to m,j}\) and \(A_{\rhat,j}^{m}\) are large in high-activity windows, so the reaction coordinate is closely tied to the coherent mean field there.

The evidence is consistent across three levels of analysis: in the representative recording, around aligned activity peaks, and after aggregation at the subject level. In all cases, elevated EHG-derived activity is associated with larger $R$, stronger reaction support, and pronounced mean-field alignment. This should be interpreted as an association within the EHG-derived electrical dynamics, not as a direct measurement of mechanical contraction force or as evidence for a causal model of parturition.

\subsection{Epileptic seizure EEG}
\label{sec:seizure_expanded}

During a seizure, scalp EEG becomes strongly organized in time and across electrodes. This makes it a test of whether an annotated clinical event coincides with changes in the reduced diagnostics \(R\) \eqref{eq:KKc_ratio_def}, \(S(\rhat)\) \eqref{eq:empirical_support}, and \(\Delta\) \eqref{eq:Delta_def}. The seizure onset is annotated independently of the diagnostics, so it serves as an external reference rather than a target the method is tuned to hit.

We use seizure recordings from the CHB-MIT dataset, sampled at \(256\,\mathrm{Hz}\). The CHB-MIT Scalp EEG Database is a public PhysioNet dataset of long-term scalp EEG recordings from pediatric subjects with intractable epilepsy, collected at Children's Hospital Boston. It contains multi-hour to multi-day recordings with expert annotations of seizure onset and offset, and is widely used as a benchmark for automated epileptic seizure detection and prediction.
We retained the EEG channels, removed dummy and non-EEG channels, normalized channel names, bandpass filtered between \(0.5\) and \(40\,\mathrm{Hz}\), applied a \(60\,\mathrm{Hz}\) notch filter, and downsampled to \(32\,\mathrm{Hz}\). Each seizure was aligned to its annotated onset,
\begin{equation}
    \tau
    =
    t-t_{\mathrm{onset}},
    \label{eq:eeg_seizure_relative_time}
\end{equation}
so \(\tau=0\) is onset. The plotted time is the window endpoint relative to onset,
\begin{equation}
    \tau_{\mathrm{end},j}
    =
    t_{\mathrm{end},j}
    -
    t_{\mathrm{onset}},
    \label{eq:eeg_window_end_relative_time}
\end{equation}
with \(t_{\mathrm{end},j}\) defined in \eqref{eq:empirical_window_end}. A diagnostic shown at \(\tau_{\mathrm{end},j}\) is thus computed from the corresponding moving window, not from a single EEG sample.

Local operators were estimated on \(40\,\mathrm{s}\) moving windows advanced in \(2\,\mathrm{s}\) steps by ridge-regularized least squares with mild isotropic shrinkage at level \(\alpha=0.05\). In each accepted window, optimization-based directed-coupling extraction (M2) and commutator-based plane extraction (M3) gave the reduced quantities \eqref{eq:empirical_reduced_quantities} and the support \eqref{eq:empirical_support}. Windows rejected by the screening rules in \cref{sec:empirical_rolling_window} were treated as missing values.

As an activity reference, we use the mean field
\begin{equation}
    m_{\mathrm{EEG}}(t)
    =
    \frac{1}{N}
    \sum_{i=1}^{N}x_i(t),
    \label{eq:eeg_mean_field}
\end{equation}
and, for the cohort summary, its smoothed absolute amplitude on the onset-aligned grid. This is a reference only; it is not part of the definition of the reduced diagnostics.

We begin with one representative seizure and then pool across the cohort. Onset-aligned curves are summarized within each patient before pooling, so that patients with many recorded seizures do not dominate the cohort median.

\begin{figure*}[!htpbt]
    \centering
    \includegraphics[width=0.95\textwidth]{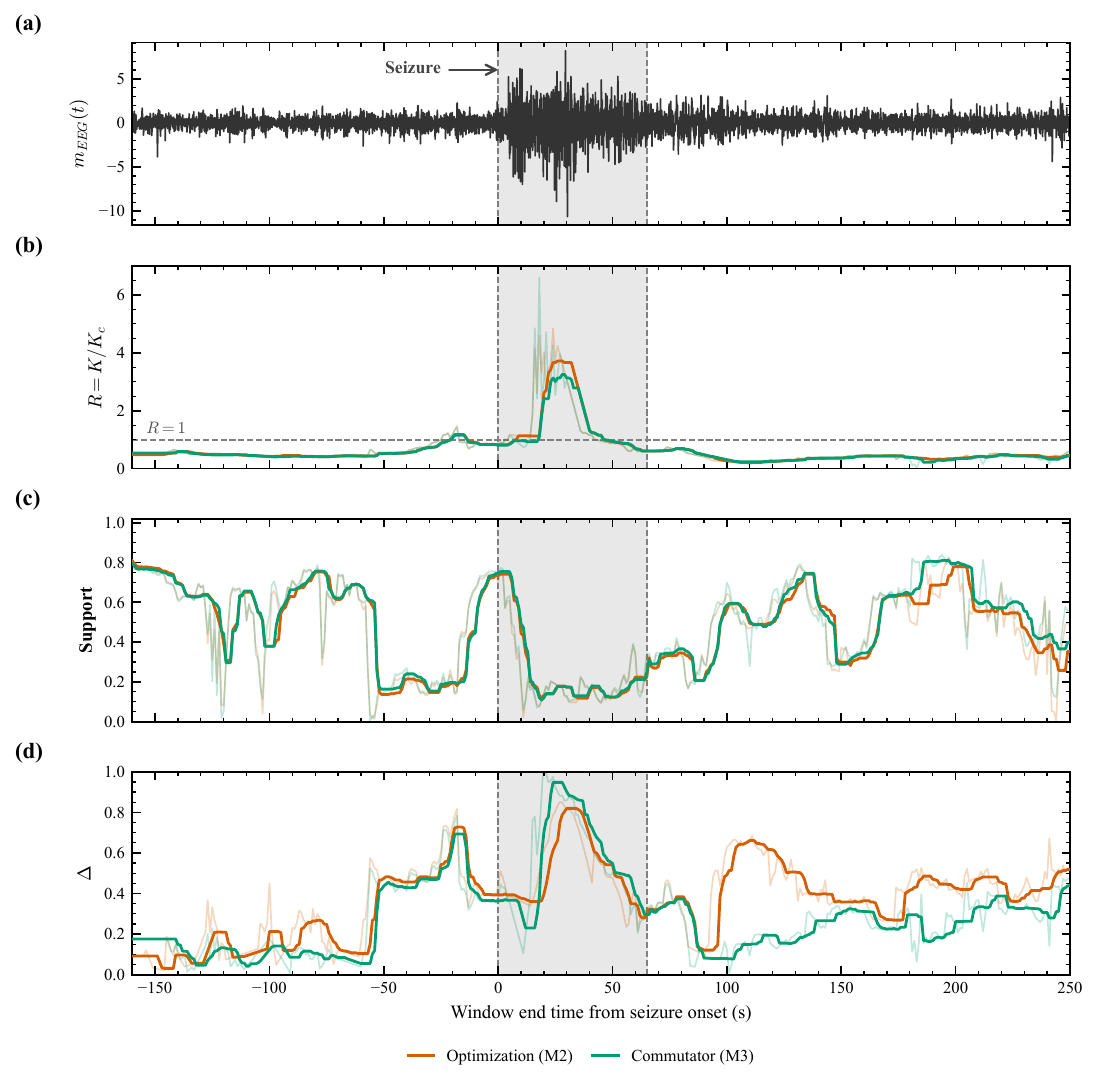}
    \caption{
    \textbf{Moving-window EEG diagnostics in one representative seizure recording.}
    The recording is CHB-MIT file \texttt{chb01\_03.edf}. The EEG was bandpass filtered between \(0.5\) and \(40\,\mathrm{Hz}\), notch filtered at \(60\,\mathrm{Hz}\), downsampled to \(32\,\mathrm{Hz}\), and analyzed with \(40\,\mathrm{s}\) moving windows advanced in \(2\,\mathrm{s}\) steps. Diagnostics are plotted at the onset-relative window endpoint \(\tau_{\mathrm{end}}\) defined in \eqref{eq:eeg_window_end_relative_time}. The shaded interval marks the annotated seizure. Orange curves denote optimization-based directed-coupling extraction (M2), and green curves denote commutator-based plane extraction (M3). Thin transparent curves show unsmoothed values, thick curves show smoothed values.
    \textbf{(a)} EEG mean field \(m_{\mathrm{EEG}}(t)\) defined in \eqref{eq:eeg_mean_field}.
    \textbf{(b)} Reduced ratio \(R\) defined in \eqref{eq:KKc_ratio_def}. The dashed horizontal line marks \(R=1\).
    \textbf{(c)} Reaction support \(S(\rhat)\) defined in \eqref{eq:empirical_support}.
    \textbf{(d)} Reduced eigenvalue splitting \(\Delta\) defined in \eqref{eq:Delta_def}.
    }
    \label{fig:seizure_representative}
\end{figure*}

\Cref{fig:seizure_representative} shows the diagnostics for one representative seizure. The mean field \eqref{eq:eeg_mean_field} in \cref{fig:seizure_representative}a increases during the annotated interval. The reduced ratio \(R\) in \cref{fig:seizure_representative}b rises sharply over the same interval for both M2 and M3. The support \(S(\rhat)\) in \cref{fig:seizure_representative}c changes during the seizure but does not fall to a single-channel response. The splitting \(\Delta\) in \cref{fig:seizure_representative}d also changes around the interval. This representative seizure therefore illustrates a clear within-recording change in the reduced geometry during the annotated ictal interval. However, because it concerns only one seizure, it should be interpreted as an illustrative example rather than as evidence for a cohort-level effect.

\begin{figure}[!htpbt]
    \centering
    \includegraphics[width=\columnwidth]{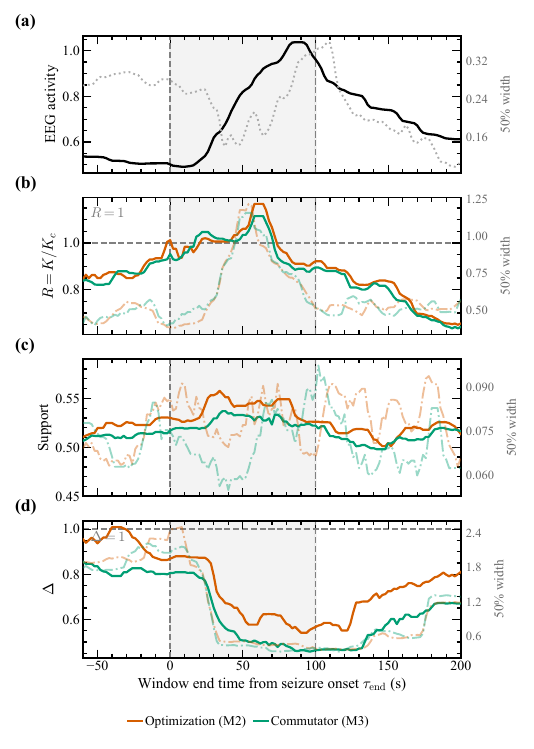}
    \caption{
    \textbf{Seizure-aligned EEG diagnostics pooled across patients.}
    The cohort analysis uses the successfully analyzed seizure recordings from the CHB-MIT dataset. Recordings were preprocessed as in \cref{fig:seizure_representative} and analyzed with \(40\,\mathrm{s}\) moving windows advanced in \(2\,\mathrm{s}\) steps. Each seizure was aligned to its annotated onset using \eqref{eq:eeg_seizure_relative_time}, and diagnostics are plotted at the onset-relative window endpoint \(\tau_{\mathrm{end}}\) defined in \eqref{eq:eeg_window_end_relative_time}. Curves were summarized within each patient before pooling. Solid curves show patient-level medians; dotted curves show the patient-level interquartile width on the right vertical axis where shown. The shaded interval marks the median annotated seizure duration. Orange curves denote optimization-based directed-coupling extraction (M2), and green curves denote commutator-based plane extraction (M3).
    \textbf{(a)} Onset-aligned EEG activity, quantified by the smoothed absolute mean field defined in \eqref{eq:eeg_mean_field}.
    \textbf{(b)} Reduced ratio \(R\) defined in \eqref{eq:KKc_ratio_def}. The dashed horizontal line marks \(R=1\).
    \textbf{(c)} Reaction support \(S(\rhat)\) defined in \eqref{eq:empirical_support}.
    \textbf{(d)} Reduced eigenvalue splitting \(\Delta\) defined in \eqref{eq:Delta_def}. The dashed horizontal line marks \(\Delta=1\).
    }
    \label{fig:seizure_dataset}
\end{figure}

\Cref{fig:seizure_dataset} pools the analysis across patients. The activity in \cref{fig:seizure_dataset}a rises after onset and stays elevated through the shaded interval marking the annotated seizures. The reduced ratio \(R\) in \cref{fig:seizure_dataset}b sits near the threshold before onset and increases modestly around the seizure. The increase is smaller than in the representative seizure, as expected after pooling across heterogeneous recordings. M2 and M3 track together in time, so the cohort pattern is not specific to one extraction method.

The support and splitting panels act as controls. The support \(S(\rhat)\) in \cref{fig:seizure_dataset}c stays roughly constant, so the response direction is not reduced to a single channel at the cohort level. The splitting \(\Delta\) in \cref{fig:seizure_dataset}d falls during the seizure. Since \(R=K/K_c(\Delta)\) couples \(K\) to \(K_c(\Delta)\), this panel is part of the reading: the cohort change in \(R\) is interpreted together with the change in \(\Delta\), not as an isolated scalar.

The representative and pooled analyses together show that the annotated seizure interval coincides with changes in \(R\) and \(\Delta\): a sharp rise in \(R\) in the representative seizure, a more moderate one after pooling. The change is not uniform across seizures, and \(R\) does not always cross \(1\).

% ------------------------------------------------------------
% ------------------------------------------------------------
\subsection{Freezing of gait}
\label{sec:fog}

Freezing of gait is an episodic motor disturbance in Parkinsonian gait in which forward progression is briefly interrupted despite the intention to walk. In inertial recordings, it usually appears as a transition from regular stride-like acceleration to a low-amplitude or irregular state. We ask whether that transition is accompanied by changes in \(R\) \eqref{eq:KKc_ratio_def}, \(S(\rhat)\)  \eqref{eq:empirical_support}, and \(\Delta\)  \eqref{eq:Delta_def}.

We analyzed recordings from the Daphnet Freezing of Gait dataset, a public benchmark dataset of wearable accelerometer signals collected from Parkinson's disease patients during walking tasks designed to elicit freezing-of-gait episodes. In our analysis, we focused on ankle accelerometry, using the annotated freezing intervals to characterize changes in the recorded motor activity.  

For the representative event, the state was formed from the three ankle acceleration channels,
\begin{equation}
    \x(t)
    =
    \big[a_x(t),a_y(t),a_z(t)\big]^\top ,
    \label{eq:fog_state}
\end{equation}
and the displayed behavioral signal is the acceleration magnitude
\begin{equation}
    \|\bm a(t)\|
    =
    \sqrt{a_x(t)^2+a_y(t)^2+a_z(t)^2}.
    \label{eq:fog_acc_mag}
\end{equation}
Local operators were estimated on \(3\,\mathrm{s}\) moving windows advanced in \(0.5\,\mathrm{s}\) steps using ridge-regularized regression. Each diagnostic is plotted at the window end time \(t_{\mathrm{end}}\) defined in \eqref{eq:empirical_window_end}. Optimization-based directed-coupling extraction (M2) and commutator-based plane extraction (M3) were applied to each accepted window to compute the reduced quantities  \eqref{eq:empirical_reduced_quantities} and the reaction support \eqref{eq:empirical_support}. Windows rejected by the screening rules in \cref{sec:empirical_rolling_window} were treated as missing values.

The representative event is \texttt{S01R01}, ankle event~03. We separate the annotated freezing episode into two parts. The onset interval is its first \(4\,\mathrm{s}\), where the gait pattern is entering the freezing state; the sustained interval is the remainder, where the ankle acceleration is strongly suppressed. The separation matters because \(R\) need not be largest when the limb is nearly motionless. The relevant question is whether \(R\) changes at the entry into freezing.

\begin{figure*}[!htpb]
    \centering
    \includegraphics[width=\textwidth]{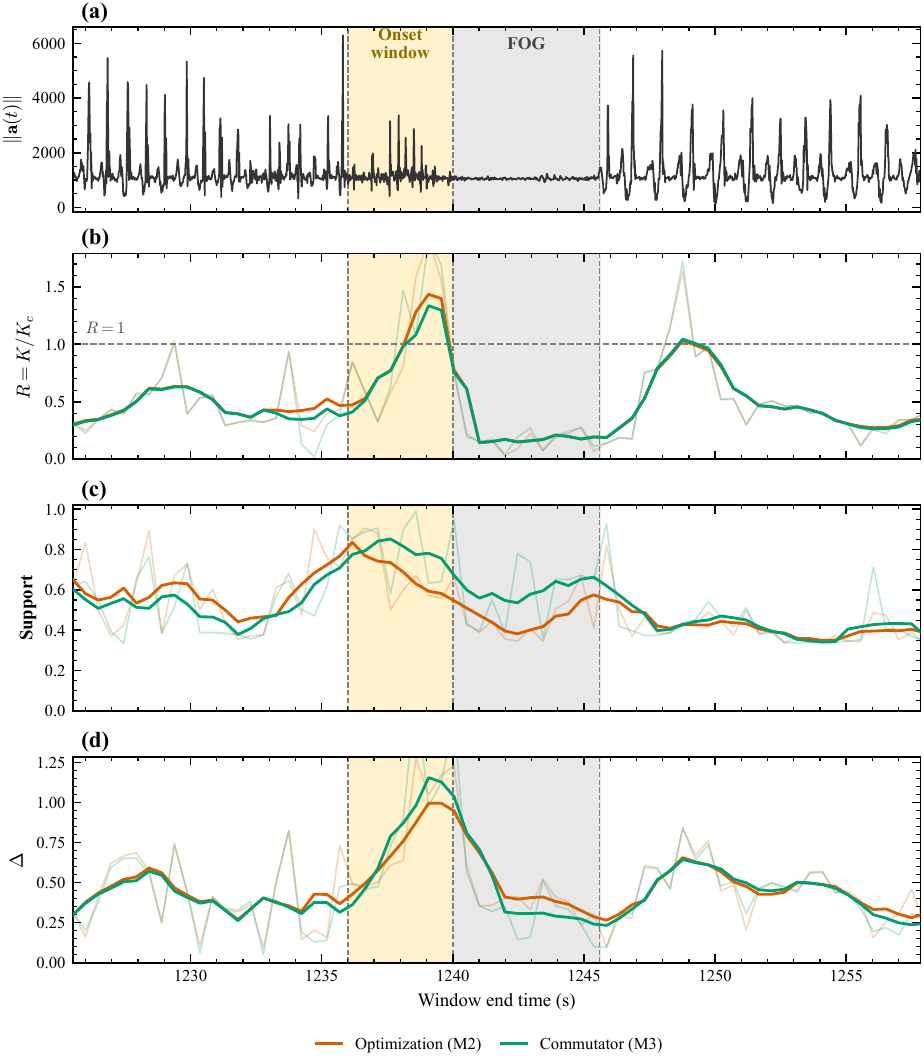}
    \caption{
    \textbf{Moving-window diagnostics in one representative freezing-of-gait (FOG) event.}
    The figure shows Daphnet recording \texttt{S01R01}, ankle event~03. The state is formed from the three ankle acceleration channels in \eqref{eq:fog_state}. Diagnostics are plotted at the window end time \(t_{\mathrm{end}}\) defined in \eqref{eq:empirical_window_end}. Orange curves denote optimization-based directed-coupling extraction (M2), and green curves denote commutator-based plane extraction (M3). Thin transparent curves show unsmoothed moving-window values, and thick curves show smoothed values. The yellow shaded interval marks the onset interval, the first \(4\,\mathrm{s}\) of the annotated freezing episode; the gray shaded interval marks the sustained interval, the remaining annotated freezing episode.
    \textbf{(a)} Ankle acceleration magnitude \(\|\bm a(t)\|\) defined in \eqref{eq:fog_acc_mag}.
    \textbf{(b)} Reduced ratio \(R\) defined in \eqref{eq:KKc_ratio_def}. The dashed horizontal line marks \(R=1\).
    \textbf{(c)} Reaction support \(S(\rhat)\) defined in \eqref{eq:empirical_support}.
    \textbf{(d)} Reduced eigenvalue splitting \(\Delta\) defined in \eqref{eq:Delta_def}.
    }
    \label{fig:fog_representative}
\end{figure*}

\Cref{fig:fog_representative} shows the behavioral signal and the moving-window diagnostics for the representative event. Before the freezing episode, the ankle acceleration \eqref{eq:fog_acc_mag} contains repeated stride-like bursts; during the sustained interval it is strongly suppressed. The reduced ratio \(R\) in \cref{fig:fog_representative}b rises across the onset interval and then falls during the low-motion sustained interval. The support \(S(\rhat)\) and \(\Delta\) also change at the onset. The largest change in \(R\) therefore occurs at the entry into freezing, not during the most motionless part of the episode.

A second rise in \(R\) can follow the freezing episode, when stride-like acceleration resumes. We read this differently from the onset rise: it likely reflects gait re-initiation rather than entry into freezing. The dataset-level comparison below therefore uses the onset interval and matched baseline windows taken outside the freezing episode.

\begin{figure}[!htpb]
    \centering
    \includegraphics[width=\columnwidth]{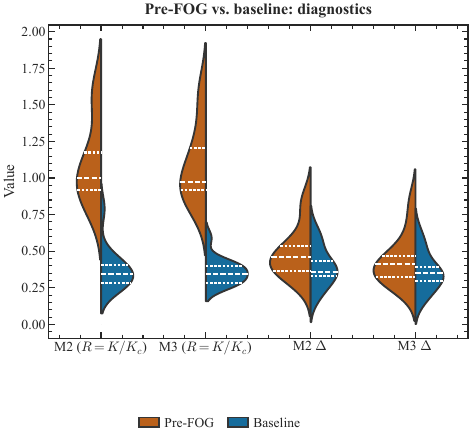}
    \caption{
    \textbf{Onset and baseline distributions of reduced diagnostics.}
    The figure summarizes freezing-of-gait events from the Daphnet recordings. For each event, the onset value is computed from the onset interval and the baseline value from matched windows outside the annotated freezing episode. Orange denotes onset windows and blue denotes baseline windows. White dashed horizontal lines mark the median and quartiles of each distribution. The first two plots show
  the reduced ratio \(R\) defined in \eqref{eq:KKc_ratio_def} obtained with the optimization-based directed-coupling extraction (M2) and commutator-based plane extraction (M3). The last two plots show the corresponding reduced eigenvalue splitting \(\Delta\) defined in \eqref{eq:Delta_def} for M2 and M3.
    }
    \label{fig:fog_dataset}
\end{figure}

\Cref{fig:fog_dataset} summarizes the same comparison across the analyzed events. The onset distributions of \(R\) sit above baseline for both M2 and M3, and the \(\Delta\) distributions show a smaller upward shift. This agrees with the representative event: \(R\) is larger in the onset interval than in matched baseline windows. The rise in \(R\) at onset is the reported effect; \(R\) is not expected to stay high through the low-motion part of the episode.

% =====================================================================
% Section 6.6  Rhythmic push-up motion  --  clean raw-first rewrite
% Logical (reader) order: analyze the signal, obtain the result, then
% ask whether the dominant cycle drives it, remove the cycle, show the
% result is unchanged. Preprocessing detail is introduced only where it
% is used (the adaptive step). New/changed prose wrapped in \newrev{}.
% Numbers are bootstrap medians, matching the bars in the figure.
% =====================================================================
\subsection{Rhythmic push-up motion}
\label{sec:pushup}

The last empirical example uses wearable inertial data recorded with an Apple Watch during rhythmic push-ups performed on two small unstable, deformable elastic fitballs, with closed fists placed parallel to the body, one on each fitball. This exercise belongs to the Logic Workout approach, which exploits controlled instability and the ``reactive falling effect'' for rehabilitation and performance enhancement \cite{sornette2025reactivefalling,sornette2026instabilityresilient} (\url{https://logicworkoutapp.com}). It differs from the three preceding datasets in that no event is externally annotated. The movement is a voluntary continuous up--down cycle on deformable supports that induce intermittent corrective adjustments; we ask whether the reduced diagnostics identify a directed input--response geometry associated with these adjustments. With no external event to align to, we separate the recording into higher- and lower-amplitude samples using its own fluctuation envelope.

The state is the three-axis user-acceleration \(\x(t)\) as in \eqref{eq:fog_state}. In the Apple Core Motion convention, gravity is already removed and acceleration is reported in units of \(g\) (about \(9.8\,\mathrm{m\,s^{-2}}\)); we subtracted the residual per-channel mean. The acceleration magnitude is defined as in \eqref{eq:fog_acc_mag}.

From the envelope of the magnitude of  \(\x(t)\), we define the \emph{high-acceleration} class as
corresponding to the upper \(15\%\) of samples and the remainder corresponds to the \emph{low-acceleration} class. For each class \(q\in\{\mathrm{low},\mathrm{high}\}\) we fit a separate local one-step operator,

\begin{equation}
    \x(t+\Delta t)\approx\Fhat_q\,\x(t),
    \qquad
    q\in\{\mathrm{low},\mathrm{high}\},
    \label{eq:pushup_regime_fit}
\end{equation}
by ridge regression, with relative ridge level \(3\times10^{-3}\) and isotropic shrinkage \(\alpha=0.04\). Applying M2 and M3 to \(\Fhat_q\) gives one reduced ratio \(R_q^{m}\) per class and method, from \eqref{eq:Delta_def}--\eqref{eq:KKc_ratio_def}. Fitting the two classes separately keeps them from being averaged into a single map.

To test whether the response direction tracks \emph{when} the high-acceleration samples occur, and not merely their size, we form the reaction coordinate
\begin{equation}
    y_r^m(t)=\left|(\rhat^{m})^\top \x(t)\right|,
    \qquad
    m\in\{\mathrm{M2},\mathrm{M3}\},
    \label{eq:pushup_reaction_projection}
\end{equation}
and quantify their temporal association by estimating the correlation between the 
reaction-coordinate envelope and the high-acceleration envelope \(e_{\mathrm{high}}(t)\),
\begin{equation}
    c_{\mathrm{high}}^m=\mathrm{corr}\!\left(e_{\mathrm{high}}(t),y_r^m(t)\right),
    \qquad m\in\{\mathrm{M2},\mathrm{M3}\}.
    \label{eq:pushup_cburst}
\end{equation}
This correlation is compared with a null obtained by circularly shifting \(y_r^m(t)\) relative to \(e_{\mathrm{high}}(t)\), which preserves its amplitude distribution and autocorrelation and destroys only its timing. With \(B=1000\) shifts and a minimum shift of \(3\,\mathrm{s}\), let
\begin{equation}
    \mu_{\mathrm{null}}^m=\frac{1}{B}\sum_{b=1}^{B}c_{\mathrm{null},b}^{m},
    \qquad
    \sigma_{\mathrm{null}}^m=\left[\frac{1}{B-1}\sum_{b=1}^{B}\left(c_{\mathrm{null},b}^{m}-\mu_{\mathrm{null}}^m\right)^2\right]^{1/2}.
    \label{eq:pushup_null_moments}
\end{equation}
The standardized score and one-sided Monte Carlo \(p\)-value are
\begin{equation}
    z^m=\frac{c_{\mathrm{high}}^m-\mu_{\mathrm{null}}^m}{\sigma_{\mathrm{null}}^m},
    \label{eq:pushup_zscore}
\end{equation}
\begin{equation}
    p^m=\frac{1+\#\{b:c_{\mathrm{null},b}^m\geq c_{\mathrm{high}}^m\}}{1+B},
    \label{eq:pushup_pvalue}
\end{equation}
so that with \(B=1000\) the smallest attainable value is \(1/1001=9.99\times10^{-4}\).

\begin{figure*}[!htpb]
    \centering
    \includegraphics[width=\textwidth]{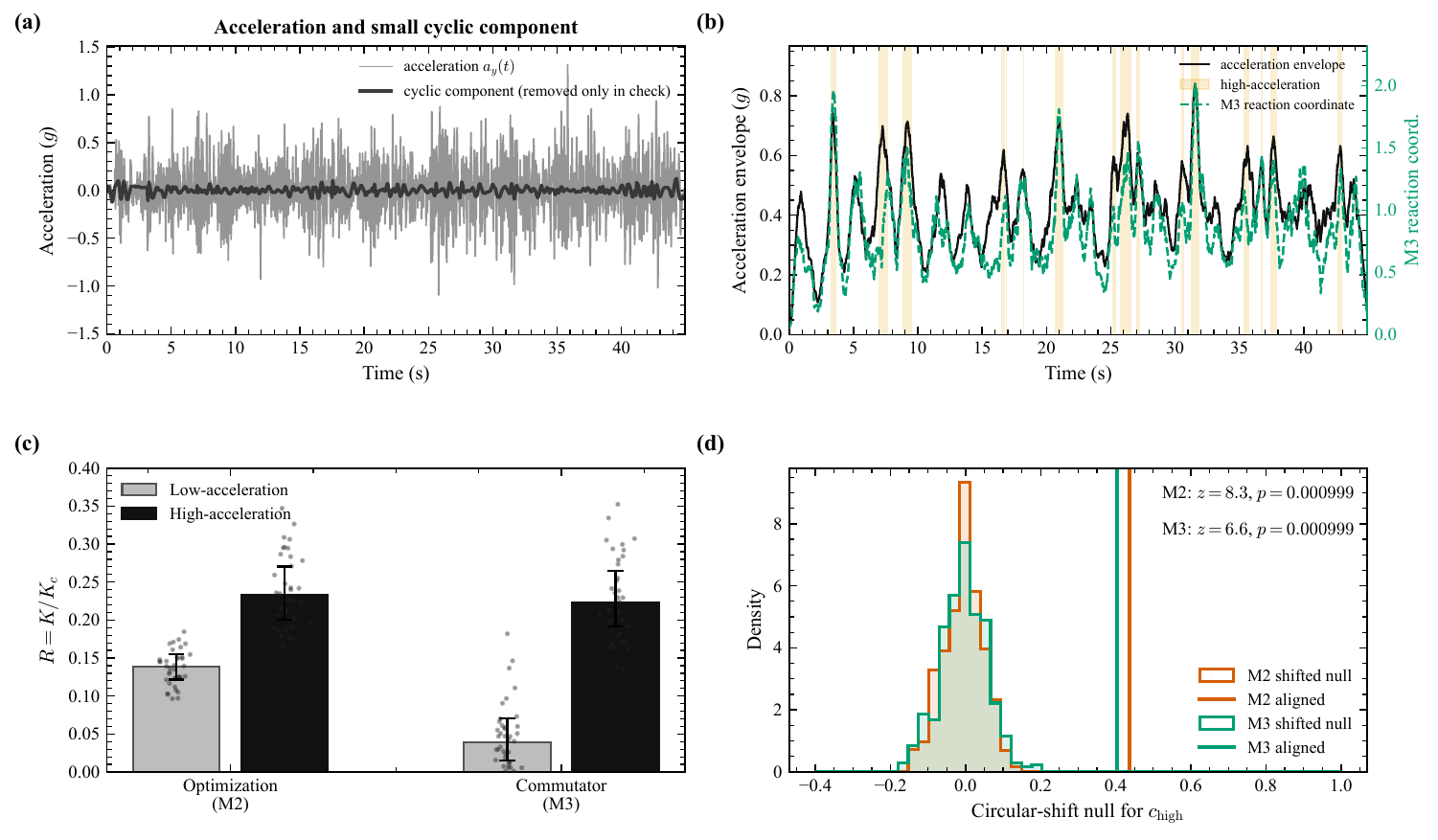}
    \caption{
    \textbf{Diagnostics for rhythmic unstable push-up motion using the LW protocol (\url{https://logicworkoutapp.com}).}
    The analyzed signal is a three-axis Apple Watch user acceleration; the channels were demeaned. The reduced diagnostics are fitted to the acceleration \(\x(t)\) \eqref{eq:fog_state}; high-acceleration samples are the upper \(15\%\) of the acceleration envelope and low-acceleration samples the remainder. Panels (a,b) show one representative axis, \(a_y(t)\), for legibility, while all three channels are used in the fit. Ridge regularization used relative ridge level \(3\times10^{-3}\) and isotropic shrinkage \(\alpha=0.04\).
    \textbf{(a)} Acceleration \(a_y(t)\) (grey) and the small slow cyclic component (dark) isolated by the phase-adaptive harmonic model of the main text, which is removed only in the robustness check described in the text.
    \textbf{(b)} Acceleration-amplitude envelope \(e_{\mathrm{high}}(t)\) with high-acceleration samples shaded, and the M3 reaction coordinate \(y_r^{\mathrm{M3}}(t)\) \eqref{eq:pushup_reaction_projection} overlaid on the right axis; these are the two slow traces whose correlation is tested in panel (d). The analyzed signal itself is the time series in panel (a).
    \textbf{(c)} Reduced ratio \(R_q^m\) for low- and high-acceleration samples, from \eqref{eq:pushup_regime_fit} and \eqref{eq:Delta_def}--\eqref{eq:KKc_ratio_def}. Bars show the bootstrap median and error bars the central \(50\%\) interval. Color encodes class (grey low-acceleration, black high-acceleration); horizontal position encodes the method (M2 left, M3 right).
    \textbf{(d)} Vertical lines mark the observed correlation \(c_{\mathrm{high}}^m\) between the high-acceleration envelope and \(y_r^m(t)\); histograms show the corresponding circular-shift null distributions (\(1000\) shifts, minimum shift \(3\,\mathrm{s}\)).
    }
    \label{fig:pushup}
\end{figure*}

\Cref{fig:pushup}a shows a representative axis of the acceleration, \(a_y(t)\), together with the slow cyclic component isolated below it; this component is small next to the broadband fluctuations. \Cref{fig:pushup}b overlays the acceleration-amplitude envelope with the M3 reaction coordinate \eqref{eq:pushup_reaction_projection}. The reaction coordinate is preferentially elevated during the shaded high-acceleration intervals, showing that its temporal modulation is aligned with the timing of these intervals. The reduced ratio is also larger for the high-acceleration class than for the low-acceleration class for both extraction methods (\Cref{fig:pushup}c). Specifically, the bootstrap median is \(R^{\mathrm{M2}}_{\mathrm{low}}\approx0.14\) and \(R^{\mathrm{M2}}_{\mathrm{high}}\approx0.23\), while \(R^{\mathrm{M3}}_{\mathrm{low}}\approx0.04\) and \(R^{\mathrm{M3}}_{\mathrm{high}}\approx0.22\); all values remain below \(R=1\). The timing test (\Cref{fig:pushup}d) gives \(c_{\mathrm{high}}^{\mathrm{M2}}\approx0.44\) and \(c_{\mathrm{high}}^{\mathrm{M3}}\approx0.40\), corresponding to \(z^{\mathrm{M2}}\approx8.3\) and \(z^{\mathrm{M3}}\approx6.6\); in both cases \(p=9.99\times10^{-4}\), the floor set by the shifts, since the observed correlation exceeds every shifted sample. The reaction coordinate is therefore temporally aligned with the high-acceleration intervals, rather than merely reflecting their larger amplitudes.

The high-acceleration class contains more of the large periodic push-up excursion than the low-acceleration class, so one may ask whether the elevated \(R\) reflects the geometry of that dominant rhythm rather than the corrective adjustments. To settle this, we remove the cycle and repeat the analysis. Because its fundamental frequency drifts across the recording, as \Cref{fig:pushup}a shows, a fixed-frequency subtraction would leave phase-dependent artifacts; we therefore fit a phase-adaptive harmonic model to each channel \(i\),
\begin{equation}
    x_i(t)=\sum_{k=1}^{K}\left[a_{ik}(t)\cos(k\phi(t))+b_{ik}(t)\sin(k\phi(t))\right]+\epsilon_i(t),
    \label{eq:pushup_adaptive_cycle_model}
\end{equation}
where \(\phi(t)\) is the instantaneous push-up phase, estimated from the first principal component of the acceleration after filtering in the \(0.2\)--\(1.2\,\mathrm{Hz}\) band. The slowly varying amplitudes \(a_{ik}(t),b_{ik}(t)\) are fit by local weighted least squares over a \(6\,\mathrm{s}\) window, with \(K=3\) harmonics. The non-cycle residual \(\epsilon(t)\) is band-limited to
\begin{equation}
    \x_{\mathrm{br}}(t)=\mathcal{B}_{1-15\,\mathrm{Hz}}\left[\epsilon(t)\right],
    \label{eq:pushup_burst_residual_state}
\end{equation}
with \(\mathcal{B}_{1-15\,\mathrm{Hz}}\) a bandpass filter, and the procedure of \eqref{eq:pushup_regime_fit}--\eqref{eq:pushup_pvalue} is repeated with \(\x_{\mathrm{br}}(t)\) in place of \(\x(t)\). The high-/low-acceleration difference is unchanged: the bootstrap median is \(R^{\mathrm{M2}}_{\mathrm{low}}\approx0.12\) and \(R^{\mathrm{M2}}_{\mathrm{high}}\approx0.24\), and \(R^{\mathrm{M3}}_{\mathrm{low}}\approx0.04\) and \(R^{\mathrm{M3}}_{\mathrm{high}}\approx0.24\), all high-acceleration values below \(1\), with the reaction coordinate again aligned to the high-acceleration intervals (\(c_{\mathrm{high}}^{\mathrm{M2}}\approx0.44\), \(c_{\mathrm{high}}^{\mathrm{M3}}\approx0.40\)); it also survives widening the residual band to \(1\)--\(25\,\mathrm{Hz}\), clipped below the Nyquist frequency. Since removing the cycle leaves the difference essentially unchanged, the excess \(R\) in the high-acceleration class is carried by the non-cycle component, not by the push-up rhythm. This is also why the cycle is kept in the EHG, seizure, and freezing-of-gait analyses: there the cyclic activity is itself the annotated event and is analyzed directly, whereas here the voluntary cycle is a confound for the corrective-adjustment question.

The result does not hinge on the \(15\%\) cutoff. Varying it from the top \(30\%\) down to the top \(9\%\) keeps \(R_{\mathrm{high}}>R_{\mathrm{low}}\) for both methods, and the separation widens monotonically as the cutoff becomes more selective (for M2, \(R_{\mathrm{high}}\) grows from \(\approx0.17\) to \(\approx0.28\)), with every value below \(1\). The effect also appears in a moving-window analysis of the kind used for the other datasets: with short windows (\(\approx1\,\mathrm{s}\), matched to the correction timescale) the local \(R\) correlates positively with the acceleration envelope (Spearman \(\rho\approx0.1\)--\(0.2\)) and is larger in high- than in low-acceleration windows, whereas the association washes out for windows \(\gtrsim2\,\mathrm{s}\) that span a full push-up cycle. This places the effect at the sub-second timescale of the reactive corrections and explains why the sample-conditioned partition, rather than the longer moving windows appropriate to the temporally extended events of the other datasets, is the natural construction here.

The point of this unstable push-up example is not that \(R\) crosses the transient-amplification threshold; it does not. It is that the high-acceleration class carries a substantially larger \(R\) than the low-acceleration class, reproducibly across the two independent extractions M2 and M3 and across the raw and cycle-removed analyses. These high-acceleration samples are therefore not merely larger-amplitude fluctuations: they are organized along a directed input--response geometry. In the Logic Workout setting, this geometry is naturally read as the footprint of reactive stabilization, in which the participant continually counters the instability of the deformable supports. Both in the raw recording and after cycle removal, the high-acceleration samples show increased reduced non-normality with a reaction coordinate locked to their timing, suggesting that the participant generates intermittent directed correction dynamics while controlling the instability of the deformable supports, characterized by their tendency to roll and to react like a spring, consistent with the reactive falling effect \cite{sornette2025reactivefalling,sornette2026instabilityresilient}. The evidence is mechanistic and suggestive, not yet proof of clinical or performance benefit.

The fact that \(R\) stays below unity here is consistent with the comparatively weak amplitude contrast of the recording. Unlike the uterine, seizure, and freezing-of-gait cases, whose signals show abrupt excursions several times larger than their background, the push-up acceleration stays within a relatively narrow range, roughly \(0.2\)--\(0.8\,g\), and reads as a continuously modulated noisy signal rather than a sharply amplified transient. Although \(R\) is inferred from the geometry of the fitted local operator rather than directly from signal amplitude, such limited contrast provides weaker evidence for a pronounced directional transfer from an input direction to a distinct response direction. The method therefore identifies some organization of the high-acceleration samples, but not an input--response geometry strong enough to cross the reduced threshold \(R=1\).

Diagnosing this subthreshold geometry is nonetheless informative, because non-normality can leave visible effects before the threshold is crossed: fluctuations become preferentially channelled along particular response directions, high-acceleration intervals align more strongly with the inferred reaction coordinate than low-acceleration intervals, and directional sensitivity can increase without producing large macroscopic excursions. The unstable push-up case is thus a useful instance of significant but sub-transient non-normal organization. 

Together, the four empirical examples then span a range of behaviours, from the clearly above-threshold episodes of the uterine, seizure, and freezing-of-gait recordings to this subthreshold regime.

% ============================================================
\section{Discussion}
\label{sec:discussion}

This work has developed a data-driven reduction framework for identifying non-normal response geometry from multivariate time series. The central object is not the full fitted operator alone, but the two-dimensional response plane \(Q=[\rhat,\nhat]\) defined in \eqref{eq:Q_def}, where \(\nhat\) is the input direction and \(\rhat\) is the response direction. The fitted dynamics are projected onto this plane through the reduced operator \(\Gam\) in \eqref{eq:reduced_operator}, and the reduced diagnostics \(\Delta\), \(\kappa_{2D}\), \(K\), and \(R\) are computed from \(\Gam\) using \eqref{eq:Delta_def}--\eqref{eq:KKc_ratio_def}. The level \(R=1\) is a reference threshold from the two-dimensional reduced calculation, not a universal clinical, behavioral, or full-system stability boundary.

The synthetic benchmarks provide the main validation. In the controlled VAR(1) family, the spectral template and orientation are fixed while the non-normal geometry is varied. As the reduced geometry becomes more non-normal, the response coordinate is increasingly expressed in the observable trajectories. This is organized by \(R\): below threshold the response stays weak, above it the reduced dynamics can produce large transient or noise-driven amplification. The mean-field correlation with the response coordinate increases with \(R\), so the inferred response direction is not only a geometric feature of the fitted operator but also appears in the observed trajectories.

The comparison of plane-extraction methods shows why the choice of reduction matters. Eigenbasis-SVD baseline extraction (M1) is a transparent reference but is fragile when the eigenvectors of the fitted operator are unstable. Optimization-based directed-coupling extraction (M2) and commutator-based plane extraction (M3) are more robust: they recover both \(R\) and the response plane in the stationary benchmark, and their agreement is informative because the two methods rest on different geometric principles. M2 maximizes directed transverse coupling; M3 uses the symmetric commutator \eqref{eq:m3_commutator} to identify a plane associated with non-normal imbalance. Agreement between them therefore indicates that the inferred plane is a property of the fitted dynamics, not an artifact of one extraction rule.

The scaling and robustness experiments clarify the target of the method. Entrywise recovery of an \(N\times N\) operator becomes harder as the dimension grows or the data shrink. Yet M2 and M3 recover the dominant response plane and the corresponding \(R\) across a broad range of dimensions, trajectory counts, and training horizons. The \(M=1\) synthetic results matter most for the empirical comparison, where each local fit comes from one observed segment. The broad singular-value benchmark shows further that the method does not assume the full operator is rank two or has only two important singular directions: the two-dimensional plane is a diagnostic projection of the dominant response geometry, not a low-rank approximation of the full operator.

The time-varying experiment extends this to non-stationary data. Moving-window inference tracks the window-matched reference \(R_{\mathrm{ref},W}\) computed from the mean true operator over the same window. This matters for empirical recordings, where a single stationary operator is rarely justified. The diagnostic reported at \(t_{\mathrm{end}}\) is a property of the data in the window \([t_{\mathrm{end}}-W,t_{\mathrm{end}})\), not of one sample. This is the convention used throughout the empirical \cref{sec:empirical}.

The empirical demonstrations show how the reduced diagnostics behave in heterogeneous recordings, read under the single standard set in \cref{sec:empirical_overview}. In EHG recordings, elevated activity coincides with larger \(R\), high reaction support, and strong alignment between the reaction coordinate and the coherent mean field, supported at three levels: one representative recording, activity-peak alignment within it, and a subject-level summary. In seizure EEG, the annotated onset interval coincides with changes in \(R\) and \(\Delta\), sharp in the representative seizure and more moderate after patient-level pooling. In freezing of gait, \(R\) rises at the onset interval and is larger there than in matched baseline windows. In unstable push-ups, \(R\) stays below \(1\), but high-acceleration samples carry a larger \(R\) than low-acceleration samples and the reaction coordinate tracks the high-acceleration intervals.

The strength of the evidence differs across these examples, and the diagnostics \(R\), \(S(\rhat)\), \(\Delta\), and the reaction-coordinate alignment measure related but distinct aspects of the reduced geometry. The EHG analysis gives the strongest, multi-level support. The seizure and freezing analyses show changes that are clear in representative recordings and more moderate in pooled summaries. The push-up example shows a directed response confined to the high-acceleration samples without crossing \(R=1\).

Several limitations remain. First, the framework depends on the quality of the local fit: short windows, low signal-to-noise ratio, ill-conditioned regression, or within-window non-stationarity can destabilize the reduced diagnostics. Ridge regularization, trace-preserving shrinkage, and screening of degenerate windows reduce this but do not remove the need for dataset-specific preprocessing and sensitivity checks. Second, the threshold \(K_c(\Delta)\) 
for the existence of transient non-normal amplifications is derived for a two-dimensional real-eigenvalue reduction; it normalizes \(K\) but is not a universal stability boundary. Third, empirical recordings provide no reference plane, so interpretation rests on the agreement of M2 and M3, temporal coherence, event alignment, support, and non-degenerate spectra. Fourth, the framework is local and linear: it can locate a directed response inside finite windows, but it does not by itself establish the nonlinear mechanism that generates an event.

Future work can extend the framework in three directions. 
Methodologically, one important direction is to derive analogous reduced diagnostics for complex-eigenvalue regimes and 
for continuous-time generators with more general spectral structures. 
Statistically, the framework should be complemented by uncertainty estimates for the inferred response plane \(Q\), 
the reduced ratio \(R\), the reaction support \(S(\hat r)\), and the eigenvalue splitting \(\Delta\), 
for example through resampling, perturbation analysis, or state-space bootstrap methods. 
Empirically, the next step is to test whether the diagnostic changes observed here persist across larger cohorts, 
alternative preprocessing pipelines, and independent datasets.  Together, these developments would move the approach beyond illustrative demonstrations toward 
a systematic inferential framework for non-normal response geometry in clinical, physiological, and behavioral time series.

\section{Conclusion}
\label{sec:conclusion}

Stable high-dimensional systems can produce large finite-time responses when their local dynamics are non-normal. Eigenvalues then set the asymptotic decay, but they do not identify the directions through which transient or noise-driven amplification occurs. This work addresses that gap with a data-driven reduction that estimates a local operator from multivariate time series, extracts a two-dimensional response plane, and computes reduced diagnostics of non-normal geometry.

The central reduced object is the projected operator \(\Gam\) in \eqref{eq:reduced_operator}, obtained from the response plane \(Q=[\rhat,\nhat]\) in \eqref{eq:Q_def}. From it, we compute the reduced eigenvalue splitting \(\Delta\), the reduced eigenvector non-orthogonality \(\kappa_{2D}\), the reduced non-normality index \(K\), and the normalized ratio \(R\) using \eqref{eq:Delta_def}--\eqref{eq:KKc_ratio_def}. The ratio \(R\) gives a common scale for comparing reduced geometry across synthetic and empirical examples; it is not a stability boundary for the full system.

The synthetic benchmarks show that the reduction recovers the dominant response plane and \(R\) from finite time series. M2 and M3 recover the reduced geometry more reliably than the eigenbasis-SVD baseline (M1), especially when the fitted operator is noisy or its eigenvectors unstable. The scaling experiments show that the response plane can be recovered even when entrywise recovery of the full operator is imperfect, and the broad singular-value benchmark shows that the two-dimensional plane is a diagnostic projection of the dominant response geometry, not a low-rank approximation of the full dynamics.

The empirical demonstrations show how the same diagnostics organize heterogeneous recordings. Elevated uterine EHG activity coincides with larger \(R\), high reaction support, and strong mean-field alignment; the annotated seizure interval coincides with changes in \(R\) and \(\Delta\); freezing of gait shows a rise in \(R\) at onset relative to baseline; and unstable push-up high-acceleration samples stay below \(R=1\) while carrying a larger \(R\) than low-acceleration samples, with the reaction coordinate tracking the high-acceleration intervals.

These are demonstrations of a diagnostic framework, not universal event detectors, and they do not replace mechanistic modeling of seizures, uterine contractions, freezing of gait, or rhythmic movement. The framework instead offers a way to identify, quantify, and visualize the low-dimensional directions through which stable local dynamics amplify fluctuations: a practical bridge from multivariate time series to interpretable non-normal response geometry.
\newpage

\appendix
% Elsevier convention: restart appendix figures/tables at A.1 (equations already do A.1)
\setcounter{figure}{0}\renewcommand{\thefigure}{\Alph{section}.\arabic{figure}}
\setcounter{table}{0}\renewcommand{\thetable}{\Alph{section}.\arabic{table}}
% ============================================================
% ============================================================
\section{Two-dimensional transient growth in a stable non-normal system}
\label{supp:two_dimensional_transient_growth}

This note gives a self-contained calculation showing how a stable two-dimensional non-normal system can exhibit finite-time amplification. The purpose is not to introduce a new estimator, but to explain the mechanism behind the reduced diagnostics used in the main text. The calculation distinguishes three different notions: spectral stability, one-step singular-value amplification, and delayed transient growth from a specified input direction.

% ------------------------------------------------------------
\subsection{Canonical discrete-time non-normal matrix}
\label{supp:canonical_matrix}

Consider the discrete-time linear system
\begin{equation}
    \x_{t+1}=F\x_t,
    \qquad
    \x_t\in \mathbb{R}^2 ,
    \label{eq:supp_discrete_system}
\end{equation}
with
\begin{equation}
    F
    =
    \begin{pmatrix}
        a & q(a-b)\\
        0 & b
    \end{pmatrix},
    \qquad
    |a|<1,\quad |b|<1,\quad q\geq 0 .
    \label{eq:supp_F_canonical}
\end{equation}
The parameter \(q\) controls the strength of the non-normal coupling. The off-diagonal entry is written as \(q(a-b)\) because this form gives a simple expression for the powers of \(F\).

The eigenvalues of \(F\) are
\begin{equation}
    \lambda_1=a,
    \qquad
    \lambda_2=b .
    \label{eq:supp_eigenvalues}
\end{equation}
Thus the system is asymptotically stable whenever \(|a|<1\) and \(|b|<1\). The matrix is non-normal unless \(q=0\) or \(a=b\). Hence the eigenvalues determine the long-time decay, while the non-normal coupling controls the possible finite-time response.

% ------------------------------------------------------------
\subsection{One-step singular-value amplification}
\label{supp:onestep_amplification}

The largest possible one-step amplification is the largest singular value of \(F\),
\begin{equation}
    \sigma_{\max}(F)
    =
    \sqrt{
    \lambda_{\max}
    \left(F^\top F\right)
    } .
    \label{eq:supp_sigma_def}
\end{equation}
For the matrix in \eqref{eq:supp_F_canonical},
\begin{equation}
    F^\top F
    =
    \begin{pmatrix}
        a^2 & aq(a-b)\\
        aq(a-b) & b^2+q^2(a-b)^2
    \end{pmatrix}.
    \label{eq:supp_FtF}
\end{equation}
Therefore
\begin{equation}
    \sigma_{\max}^2(F)
    =
    \frac{
        \tau+
        \sqrt{\tau^2-4a^2b^2}
    }{2},
    \qquad
    \tau=a^2+b^2+q^2(a-b)^2 .
    \label{eq:supp_sigmax_formula}
\end{equation}
One-step growth occurs if and only if
\begin{equation}
    \sigma_{\max}(F)>1 .
    \label{eq:supp_onestep_growth_condition}
\end{equation}
Solving \(\sigma_{\max}(F)=1\) gives
\begin{equation}
    q_{c,1}
    =
    \frac{
        \sqrt{(1-a^2)(1-b^2)}
    }{
        |a-b|
    } .
    \label{eq:supp_qc_onestep}
\end{equation}
Thus a stable matrix can amplify some initial condition after one step if
\begin{equation}
    q>q_{c,1}.
    \label{eq:supp_q_onestep_condition}
\end{equation}
The initial condition that maximizes one-step amplification  is the dominant right singular vector of \(F\), not an eigenvector. This is the first reason why transient amplification cannot be inferred from eigenvalues alone.

% ------------------------------------------------------------
\subsection{Exact powers and delayed transient growth}
\label{supp:exact_powers}

The powers of the canonical matrix are
\begin{equation}
    F^t
    =
    \begin{pmatrix}
        a^t & q(a^t-b^t)\\
        0 & b^t
    \end{pmatrix},
    \qquad
    t=0,1,2,\ldots .
    \label{eq:supp_F_power}
\end{equation}
Starting from the input coordinate
\begin{equation}
    e_n=
    \begin{pmatrix}
        0\\
        1
    \end{pmatrix},
    \label{eq:supp_en}
\end{equation}
we obtain
\begin{equation}
    F^t e_n
    =
    \begin{pmatrix}
        q(a^t-b^t)\\
        b^t
    \end{pmatrix}.
    \label{eq:supp_Ft_en}
\end{equation}
The second component decays as \(b^t\), but during this decay it feeds the first component through the non-normal coupling \(q(a^t-b^t)\). The response norm is
\begin{equation}
    \|F^t e_n\|_2
    =
    \sqrt{
        q^2(a^t-b^t)^2+b^{2t}
    } .
    \label{eq:supp_growth_exact}
\end{equation}
For large \(q\), the first component dominates during the transient phase, and
\begin{equation}
    \|F^t e_n\|_2
    \simeq
    q\,|a^t-b^t| .
    \label{eq:supp_growth_envelope}
\end{equation}
This is the simplest expression of the mechanism: a stable input coordinate can be redirected into a response coordinate before both components decay asymptotically.

% ------------------------------------------------------------
\subsection{Peak time for the transient response}
\label{supp:peak_time}

Assume for clarity that
\begin{equation}
    0<b<a<1 .
    \label{eq:supp_ab_positive_ordered}
\end{equation}
The large-\(q\) transient envelope is proportional to
\begin{equation}
    g(t)=a^t-b^t .
    \label{eq:supp_gt}
\end{equation}
Treating \(t\) as continuous, the maximum satisfies
\begin{equation}
    \frac{d}{dt}g(t)
    =
    a^t\log a-b^t\log b=0 .
    \label{eq:supp_peak_condition}
\end{equation}
Therefore
\begin{equation}
    \left(\frac{a}{b}\right)^{t_\ast}
    =
    \frac{\log b}{\log a},
    \label{eq:supp_peak_ratio}
\end{equation}
and
\begin{equation}
    t_\ast
    =
    \frac{
        \log\!\left(\log b/\log a\right)
    }{
        \log(a/b)
    } .
    \label{eq:supp_peak_time}
\end{equation}
Both \(\log a\) and \(\log b\) are negative, so the ratio \(\log b/\log a\) is positive. If \(a\) and \(b\) are close, the two stable decay rates compete over a longer time and the peak is delayed. If \(b\) is much smaller than \(a\), the peak occurs earlier.

The corresponding continuous-time peak height of the large-\(q\) envelope is
\begin{equation}
    g_\ast
    =
    a^{t_\ast}-b^{t_\ast}.
    \label{eq:supp_gstar_def}
\end{equation}
Using \eqref{eq:supp_peak_ratio},
\begin{equation}
    b^{t_\ast}
    =
    a^{t_\ast}\frac{\log a}{\log b},
    \label{eq:supp_bstar_astar}
\end{equation}
so that
\begin{equation}
    g_\ast
    =
    a^{t_\ast}
    \left(
        1-\frac{\log a}{\log b}
    \right).
    \label{eq:supp_gstar}
\end{equation}
For the exact discrete-time system, the maximum is obtained by checking the nearest integers to \(t_\ast\),
\begin{equation}
    t_{\mathrm{peak}}
    \in
    \left\{
        \lfloor t_\ast\rfloor,\,
        \lceil t_\ast\rceil
    \right\},
    \label{eq:supp_discrete_peak}
\end{equation}
and choosing the integer that maximizes \(\|F^t e_n\|_2\).

% ------------------------------------------------------------
\subsection{Exact delayed-peak threshold for the canonical matrix}
\label{supp:peak_transient_threshold}

One-step amplification and delayed transient amplification are related but not identical. The one-step threshold \(q_{c,1}\) in \eqref{eq:supp_qc_onestep} asks whether some optimally chosen initial condition grows after one iterate. A different question is whether the specified input coordinate \(e_n\) produces a delayed response larger than its initial size.

For the exact response in \eqref{eq:supp_Ft_en}, peak amplification from \(e_n\) occurs if there exists an integer \(t\geq 1\) such that
\begin{equation}
    q^2(a^t-b^t)^2+b^{2t}>1 .
    \label{eq:supp_peak_growth_exact_condition}
\end{equation}
Equivalently, the exact discrete delayed-peak threshold is
\begin{equation}
    q_{c,\mathrm{peak}}^{\mathrm{disc}}
    =
    \min_{t\geq 1}
    \frac{
        \sqrt{1-b^{2t}}
    }{
        |a^t-b^t|
    } .
    \label{eq:supp_qc_peak_discrete}
\end{equation}
In the strongly non-normal regime, the \(b^t\) term in \eqref{eq:supp_growth_exact} is subdominant near the peak, and a useful approximation is
\begin{equation}
    q_{c,\mathrm{peak}}
    \simeq
    \frac{1}{
        \max_{t\geq 0}|a^t-b^t|
    }
    =
    \frac{1}{g_\ast}.
    \label{eq:supp_qc_peak_approx}
\end{equation}
This threshold depends on the specified eigenvalue pair \((a,b)\). It is therefore a per-matrix threshold, not the scale-normalized threshold used for comparing different inferred reduced planes.

% ------------------------------------------------------------
\subsection{Illustration of delayed transient amplification}
\label{supp:delayed_transient_figure}

The calculation above is a finite-dimensional version of the classical nonmodal transient-growth mechanism: eigenvalues determine asymptotic decay, while singular vectors and finite-time propagators determine the largest finite-time response \cite{trefethen1993hydrodynamic,trefethen2005spectra}. In stochastic systems, the same geometry can amplify noise-driven fluctuations even when the underlying dynamics remain linearly stable \cite{farrell1994variance}. The canonical matrix in \eqref{eq:supp_F_canonical} makes this distinction explicit.

\begin{figure*}[t]
    \centering
    \includegraphics[width=\textwidth]{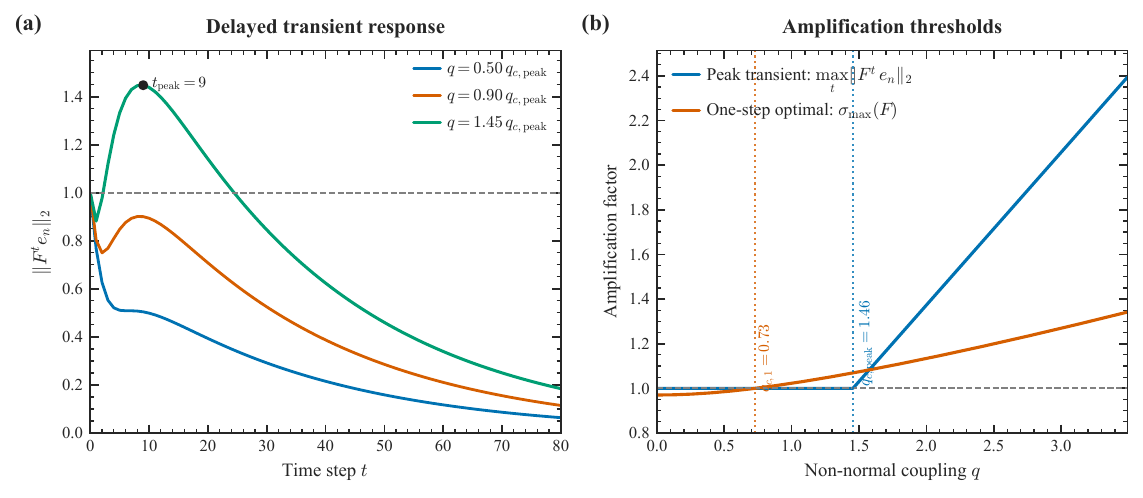}
    \caption{
    \textbf{Finite-time amplification in a stable two-dimensional non-normal system.}
    The matrix is
    \(F=\left(\begin{smallmatrix}a&q(a-b)\\0&b\end{smallmatrix}\right)\),
    with \(a=0.97\), \(b=0.75\), and \(0<b<a<1\). The eigenvalues are
    \(\lambda_1=a\) and \(\lambda_2=b\), so both lie inside the unit disk. The input coordinate is \(e_n=(0,1)^\top\).
    \textbf{(a)} Response norm \(\|F^t e_n\|_2\) for
    \(q=0.50\,q_{c,\mathrm{peak}}\),
    \(q=0.90\,q_{c,\mathrm{peak}}\), and
    \(q=1.45\,q_{c,\mathrm{peak}}\). For these parameters,
    \(q_{c,\mathrm{peak}}\simeq1.46\), and the displayed super-threshold response peaks near \(t_{\mathrm{peak}}=9\).
    \textbf{(b)} Amplification factors as functions of the non-normal coupling \(q\). The blue curve shows the delayed peak response
    \(G_{\mathrm{peak}}(q)=\max_t\|F^t e_n\|_2\). The orange curve shows the one-step optimal amplification
    \(\sigma_{\max}(F)\). The blue vertical dotted line marks \(q_{c,\mathrm{peak}}\simeq1.46\), and the orange vertical dotted line marks the one-step threshold \(q_{c,1}\simeq0.73\). The grey horizontal dashed line marks amplification factor \(1\).
    }
    \label{fig:supp_2d_transient_growth}
\end{figure*}

\Cref{fig:supp_2d_transient_growth}a shows the delayed response from the input coordinate \(e_n\). For \(q<q_{c,\mathrm{peak}}\), the response does not exceed its initial norm. For \(q>q_{c,\mathrm{peak}}\), the response grows to a finite-time peak and then decays. This panel illustrates the delayed transfer from the input coordinate into the response coordinate, even though both eigenvalues lie inside the unit disk.

\Cref{fig:supp_2d_transient_growth}b compares two amplification thresholds for the same stable matrix family. The blue curve is the delayed peak response
\begin{equation}
    G_{\mathrm{peak}}(q)
    =
    \max_{t\geq 0}
    \|F^t e_n\|_2 ,
    \label{eq:supp_peak_gain_q}
\end{equation}
where the initial direction is fixed to be \(e_n\). The orange curve is the one-step optimal gain
\begin{equation}
    \sigma_{\max}(F)
    =
    \max_{\|\x_0\|_2=1}
    \|F\x_0\|_2 .
    \label{eq:supp_sigma_max_interpretation}
\end{equation}
The two thresholds differ because they answer different questions. The one-step threshold \(q_{c,1}\) asks whether some optimally chosen initial condition grows after one iterate. The delayed-peak threshold \(q_{c,\mathrm{peak}}\) asks whether the specified input coordinate \(e_n\) produces a response above its initial norm at a later time.

This distinction also clarifies the empirical interpretation in the main text. A reaction coordinate can align with observed bursts even when \(R<1\). By contrast, \(R>1\) indicates that the inferred reduced plane lies above the scale-normalized reduced threshold used in the two-dimensional diagnostic.

% ============================================================
\section{Basis-invariant form of the reduced threshold}
\label{supp:two_dimensional_threshold}

\ref{supp:two_dimensional_transient_growth} used the canonical matrix
\begin{equation}
    F
    =
    \begin{pmatrix}
        a & q(a-b)\\
        0 & b
    \end{pmatrix},
    \qquad
    |a|<1,\quad |b|<1,
    \label{eq:supp2_F_canonical}
\end{equation}
to show how a stable two-dimensional system can produce finite-time amplification. Here we connect the coordinate-dependent parameter \(q\) to the basis-invariant reduced quantities used in the main text.

The eigenvalues of $F$ \eqref{eq:supp2_F_canonical} are
\begin{equation}
    \lambda_1=a,
    \qquad
    \lambda_2=b .
    \label{eq:supp2_eigenvalues}
\end{equation}
The reduced eigenvalue splitting is
\begin{equation}
    \Delta
    =
    \left|
    \frac{\lambda_1-\lambda_2}
         {\lambda_1+\lambda_2}
    \right|
    =
    \left|
    \frac{a-b}{a+b}
    \right| .
    \label{eq:supp2_Delta}
\end{equation}
Thus \(\Delta\) measures the relative separation of the two reduced eigenvalues.

For \(a\neq b\), the right eigenvectors may be chosen as
\begin{equation}
    v_1
    =
    \begin{pmatrix}
        1\\
        0
    \end{pmatrix},
    \qquad
    v_2
    =
    \begin{pmatrix}
        -q\\
        1
    \end{pmatrix}.
    \label{eq:supp2_v12}
\end{equation}
After normalization,
\begin{equation}
    p_1
    =
    \begin{pmatrix}
        1\\
        0
    \end{pmatrix},
    \qquad
    p_2
    =
    \frac{1}{\sqrt{1+q^2}}
    \begin{pmatrix}
        -q\\
        1
    \end{pmatrix}.
    \label{eq:supp2_unit_eigenvectors}
\end{equation}
Their absolute inner product is
\begin{equation}
    |\langle p_1,p_2\rangle|
    =
    \frac{q}{\sqrt{1+q^2}} .
    \label{eq:supp2_inner_product}
\end{equation}
Substitution into the reduced eigenvector-conditioning diagnostic gives
\begin{align}
    \kappa_{2D}
    &=
    \sqrt{
    \frac{1+|\langle p_1,p_2\rangle|}
         {1-|\langle p_1,p_2\rangle|}
    } \nonumber\\
    &=
    \sqrt{
    \frac{1+q/\sqrt{1+q^2}}
         {1-q/\sqrt{1+q^2}}
    }
    =
    \sqrt{1+q^2}+q .
    \label{eq:supp2_kappa2D_q}
\end{align}
Therefore
\begin{equation}
    \kappa_{2D}^{-1}
    =
    \sqrt{1+q^2}-q,
    \label{eq:supp2_kappa_inverse}
\end{equation}
and the reduced non-normality index becomes
\begin{equation}
    K
    =
    \frac{\kappa_{2D}-\kappa_{2D}^{-1}}{2}
    =
    q .
    \label{eq:supp2_K_equals_q}
\end{equation}
Thus, in the canonical triangular model, the coordinate coupling \(q\) is exactly equal to the basis-invariant reduced non-normality index \(K\).

The exact delayed-peak threshold in \eqref{eq:supp_qc_peak_discrete} depends on the absolute eigenvalue pair \((a,b)\). This dependence is appropriate when the full two-dimensional matrix is specified. The main text, however, compares reduced planes across windows, datasets, and fitted operators. For this purpose, we use a scale-normalized threshold that depends only on the reduced eigenvalue splitting \(\Delta\):
\begin{equation}
    K_c^2(\Delta)
    =
    \frac{\sqrt{1-\Delta^2}}
         {1-\sqrt{1-\Delta^2}},
    \qquad
    0\leq \Delta<1 .
    \label{eq:supp2_Kc_squared}
\end{equation}
Equivalently,
\begin{equation}
    K_c(\Delta)
    =
    \left[
    \frac{\sqrt{1-\Delta^2}}
         {1-\sqrt{1-\Delta^2}}
    \right]^{1/2}.
    \label{eq:supp2_Kc_final}
\end{equation}
This threshold is not the same object as \(q_{c,\mathrm{peak}}^{\mathrm{disc}}\). The latter is an exact delayed-peak threshold for a specified matrix and a specified input direction. In contrast, $K_c(\Delta)$ is a scale-normalized reduced threshold expressed in the basis-invariant diagnostics \(\Delta\) and \(K\).

The limiting behavior of \(K_c(\Delta)\) is consistent with the role of spectral splitting. As \(\Delta\to0\), the two reduced eigenvalues become nearly equal and
\begin{equation}
    K_c(\Delta)\to\infty .
    \label{eq:supp2_Kc_limit_zero}
\end{equation}
In this limit, a larger amplitude of reduced eigenvector non-orthogonality is required to obtain the same scale-normalized amplification level. As \(\Delta\to1\),
\begin{equation}
    K_c(\Delta)\to0 .
    \label{eq:supp2_Kc_limit_one}
\end{equation}
In this limit, the two reduced eigenvalues are strongly separated and less non-orthogonality is needed to reach the reduced threshold.

The normalized reduced non-normality ratio used in the main text is
\begin{equation}
    R
    =
    \frac{K}{K_c(\Delta)} .
    \label{eq:supp2_R}
\end{equation}
For the canonical triangular model, since \(K=q\), this can be written as
\begin{equation}
    R
    =
    \frac{q}{K_c(\Delta)} .
    \label{eq:supp2_R_q}
\end{equation}
Values \(R<1\) indicate that the inferred reduced geometry lies below the scale-normalized reduced threshold, while values \(R>1\) indicate that it lies above that threshold. This statement concerns the inferred two-dimensional reduced geometry. It does not imply asymptotic instability of the full system.

The distinction between \(q_{c,\mathrm{peak}}^{\mathrm{disc}}\) and \(K_c(\Delta)\) is essential. The exact threshold \(q_{c,\mathrm{peak}}^{\mathrm{disc}}\) is tied to a specified eigenvalue pair and a specified initial direction. The reduced threshold \(K_c(\Delta)\) is used to compare inferred planes through \(\Delta\), \(\kappa_{2D}\), and \(K\). Consequently, \(R=K/K_c(\Delta)\) should be interpreted as a reduced diagnostic of the inferred plane, not as a universal full-system stability boundary.

\section*{CRediT authorship contribution statement}
\textbf{V.R. Saiprasad:} Software, Investigation, Formal analysis, Validation, Writing -- original draft, Writing -- review \& editing. \textbf{V. Troude:} Conceptualization, Methodology, Writing -- original draft, Writing -- review \& editing. \textbf{D. Sornette:} Conceptualization, Supervision, Funding acquisition, Writing -- review \& editing. V.R. Saiprasad and V. Troude contributed equally to this work.

\section*{Declaration of competing interest}

The authors declare that they have no known competing financial interests or personal relationships that could have appeared to influence the work reported in this paper.

\section*{Funding}

D.S. was partially supported by the National Natural Science Foundation of China (Grant No. T2350710802 and No. U2039202), the Shenzhen Science and Technology Innovation Commission (Grants No. GJHZ20210705141805017 and No. K23405006), and the Center for Computational Science and Engineering at the Southern University of Science and Technology.

\section*{Data and code availability}

The non-normal directional response inference method is released as an open-source, reusable calibration package, provided together with the code that reproduces the analyses and synthetic benchmarks of this paper and a per-dataset reproducibility guide in a single public repository \cite{nndr_package}, which also contains the push-up inertial recording analysed here. The three public datasets are available from their original sources: the multichannel electrohysterogram recordings and the CHB-MIT Scalp EEG Database through PhysioNet \cite{goldberger2000physiobank,alexandersson2015ehg}, and the Daphnet Freezing-of-Gait dataset \cite{bloem2004fog}.

\bibliographystyle{elsarticle-num}
\bibliography{references}

\end{document}